\documentclass[draft]{agujournal2018}
\usepackage{apacite}
\usepackage{url} 
\usepackage{lineno}
\usepackage{caption}
\usepackage{graphicx,setspace}
\usepackage{bm}
\usepackage{amsmath}
\usepackage{amssymb}
\usepackage[version=4]{mhchem}
\usepackage{siunitx}
\DeclareSIUnit\year{yr}
\DeclareSIUnit\carbon{C}

\draftfalse
\journalname{Journal of Advances in Modeling Earth Systems (JAMES)}

\begin{document}

\title{Effects of Langmuir Turbulence on Upper Ocean Carbonate Chemistry}
\authors{K. M. Smith\affil{1}, P. E. Hamlington\affil{1}, K. E. Niemeyer\affil{2}, B. Fox-Kemper\affil{3}, \\ N. S. Lovenduski\affil{4,5}}
\affiliation{1}{Department of Mechanical Engineering, University of Colorado, Boulder, CO}
\affiliation{2}{School of Mechanical, Industrial, and Manufacturing Engineering, Oregon State University, Corvallis, OR}
\affiliation{3}{Department of Earth, Environmental, and Planetary Sciences, Brown University, Providence, RI}
\affiliation{4}{Department of Atmospheric and Oceanic Sciences, University of Colorado, Boulder, CO}
\affiliation{5}{Institute of Arctic and Alpine Research, University of Colorado, Boulder, CO}
\correspondingauthor{Peter Hamlington}{peh@colorado.edu}

\begin{keypoints}
\item Detailed carbonate chemistry is solved in large eddy simulations of upper ocean turbulence.
\item Langmuir turbulence increases the air-sea flux of CO$_2$, resulting in increased dissolved inorganic carbon.
\item Equilibrium chemistry leads to over-predicted fluxes of CO$_2$ into the upper ocean.
\end{keypoints}

\begin{abstract}
Effects of wave-driven Langmuir turbulence on the air-sea flux of carbon dioxide (\ce{CO2}) are examined using large eddy simulations featuring actively reacting carbonate chemistry in the ocean mixed layer at small scales. Four strengths of Langmuir turbulence are examined with three types of carbonate chemistry: time-dependent, instantaneous equilibrium chemistry, and no reactions. The time-dependent model is obtained by reducing a detailed eight-species chemical mechanism using computational singular perturbation analysis, resulting in a quasi-steady-state approximation for hydrogen ion (\ce{H+}), i.e., fixed pH. The reduced mechanism is then integrated in two half-time steps before and after the advection solve using a Runge--Kutta--Chebyshev scheme that is robust for stiff systems of differential equations. The simulations show that, as the strength of Langmuir turbulence increases, \ce{CO2} fluxes are enhanced by rapid overturning of the near-surface layer, which rivals the removal rate of \ce{CO2} by time-dependent reactions. Equilibrium chemistry and non-reactive models are found to bring more and less carbon, respectively, into the ocean as compared to the more realistic time-dependent model. These results have implications for Earth system models that either neglect Langmuir turbulence or use equilibrium, instead of time-dependent, chemical mechanisms.
\end{abstract}

\section{Introduction}
The ocean is a critical component of the global carbon cycle, presently holding 60 times more carbon than the pre-industrial atmosphere \citep{ciais2013}. Changes in ocean carbon storage also affect atmospheric carbon dioxide (\ce{CO2}), thereby impacting the climate system. From a dynamical perspective, ocean carbon uptake is intricately linked to physical circulations \citep{graven2012}, and recent research emphasizes the role of large-scale fluid advective processes in transporting carbon across the base of the mixed layer \citep{levy2013}. However, few studies have explored the role of small-scale turbulent circulations in ocean carbon uptake, resulting in continued uncertainty regarding parameterizations of air-sea \ce{CO2} fluxes in Earth system models (ESMs). The present study examines the ocean carbon cycle at small scales using large eddy simulations (LES) to model reactive carbonate species evolving in the presence of realistic mixed-layer turbulence.

Prior research on reacting flows (e.g., \citep{hamlington2011}) has shown that the strongest interactions between reactions and turbulence occur when chemical and mixing timescales are within an order of magnitude, as is the case for carbonate chemistry in the oceanic mixed layer. In the ocean surface boundary layer, cooling-driven surface convection, wind-driven shear turbulence, and wave-driven Langmuir turbulence \citep{langmuir1938} occur at time scales of roughly \SIrange{1}{100}{\minute}, while wave periods and breaking occur at roughly \SIrange{1}{10}{\second}. Once \ce{CO2} is transferred across the air-sea interface, it reacts with seawater to produce bicarbonate (\ce{HCO3^-}) and carbonate (\ce{CO3^2-}) in a series of reactions whose rate-limiting steps have time scales of roughly \SI{1}{\minute} \citep{zeebe2001}. As a result, the timescales of small-scale ocean turbulence and carbonate chemistry can be of the same order, leading to strong coupling between flow physics and reactions. Larger-scale turbulent processes, such as mixing by meso- and submeso-scale eddies, have much longer timescales and are not likely to interact strongly with carbonate chemistry reactions, although they may link in a similar way to the biological carbon cycle, which also has long timescales. Although there are linkages between inorganic and biological carbon cycles that can cause slow variations in chemical composition, chemical reactions themselves remain fast and are coupled most strongly to correspondingly fast turbulent mixing.  
     
Despite the timescale matching between carbonate chemistry and turbulence, however, the present study is the first to simultaneously solve chemical and fluid flow equations in a coupled fashion. Previous studies have shown that Langmuir turbulence produces spatial heterogeneity (or ``patchiness'') at the ocean surface through the aggregation of buoyant tracers such as debris, plankton, nutrients, or oil within the convergence zones of its counter rotating cells \citep{langmuir1938,stommel1949,smayda1970,powell1975,barstow1982,thorpe2000,thorpe2009,lewis2005,qiao2009,smith2016,suzuki2016}. Due to enhanced mixing, Langmuir turbulence also increases the vertical extent over which tracers are distributed \citep{woodcock1950,woodcock1993,shoener1970,johnson1977,ledbetter1979,buranathanitt1982,pinelalloul1995,solow1995,gallager1996,yool1998,bees1998,mcwilliams2000}. The effects of surface waves on sea-surface chemistry have also been examined \citep{sutcliffe1963,williams1967,parsons1973,eisenreich1978,dierssen2009}, largely focusing on the increase in aggregation of nutrients due to windrows and the precipitation of organic particles due to bubble injection from surface wave breaking. In each of these cases, however, reactions were assumed to be sufficiently fast or slow in comparison to the dominant mixing process and, as a result, significant reaction-flow couplings were neglected.
     
Although some studies have examined time-dependent carbonate chemistry within the ocean, primarily focusing on enhancement of air-sea fluxes and uptake of carbon by individual phytoplankton cells, these studies have largely assumed that the flow is laminar, quiescent, or well-mixed by small-scale turbulence \citep{quinn1971,johnson1982,williams1983,jahne1995,wolf-gladrow1999,schulz2006,zeebe2007,schulz2009,guo2011,uchikawa2012}. Many studies have also examined the enhancement of \ce{CO2} exchange rates across the air-sea interface as a function of wind-driven turbulence, wave breaking, and bubble injection; however, no studies have specifically included a time-dependent \ce{CO2} hydration mechanism \citep{bolin1960,hoover1969,pankow1982,goldman1982,asher1986,woolf1993,jahne1995,farmer1995,asher1998,boutin1999,jacobs2002,thorpe2003,kuss2004}. 
     
The primary goals of this paper are to examine the effects of Langmuir turbulence on carbonate chemistry by simultaneously considering the time-dependent nature of both processes and to exemplify a coupled chemistry-physics modeling system capable of carrying out this examination. Specifically, this study seeks to determine how Langmuir turbulence affects the amount of dissolved inorganic carbon (DIC) in the oceanic mixed layer. As a secondary objective, this study examines how chemical model fidelity affects predictions of DIC in the mixed layer. These objectives are addressed using LES of reactive carbonate species at small scales for different strengths of Langmuir turbulence and different chemical models, including time-dependent and equilibrium models. The simulations are enabled by the development of a new reduced mechanism for carbonate chemistry, as well as the implementation of a Runge--Kutta--Chebyshev numerical integrator to handle the stiffness of the governing equations. Without these developments, the computational cost of the LES is prohibitive.

In the following, details of the numerical simulations are provided in Section \ref{sec:sims}, including the development of the reduced carbonate chemistry model. Section \ref{results} outlines the simulation results and Section \ref{sec:discuss} discusses the implications of these results for ESMs, as well as how the present observations might vary for different ocean conditions. Conclusions and directions for future research are provided at the end.

\section{Description of Numerical Simulations\label{sec:sims}}

\subsection{Governing Equations and Solver}
The governing equations solved in the simulations are the wave-averaged Boussinesq equations \citep{suzuki2016} with additional transport equations for reactive species concentrations (termed ``tracers'' in the following) \citep{smith2016}, namely
\begin{linenomath*}
\begin{eqnarray}\label{mom_eq} 
\frac{D {\bf u}}{Dt}	&=& - \nabla p -{\bf f}_\mathrm{c} \times {\bf u}_\mathrm{L} - u_{\mathrm{L},j} \nabla u_{\mathrm{s},j} + b \hat{{\bf z}}+ \textbf{SGS}_u\,,\\
\frac{D b}{Dt} 		&=& \mathrm{SGS}_b\,,\label{buoy}\\
\frac{D {\bf c}}{Dt}	&=& {\bf S} + {\bf SGS}_c\,, \label{concentration_cont} \\
\nabla \cdot {\bf u} 	&=& 0\,,\label{continuity}
\end{eqnarray}
\end{linenomath*}
where $D/Dt \equiv \partial/\partial t + ({\bf u}_\mathrm{L} \cdot \nabla)$ is the material derivative, ${\bf u}_\mathrm{L} \equiv {\bf u} + {\bf u}_\mathrm{s}$ is the Lagrangian velocity, ${\bf u}$ is the Eulerian velocity averaged over surface gravity waves, ${\bf u}_\mathrm{s}$ is the Stokes drift velocity created by surface gravity waves, $p$ is the pressure normalized by a reference density $\rho_0$, ${\bf f}_\mathrm{c}$ is the Coriolis parameter, and $b$ is the buoyancy. Buoyancy and density $\rho$ are related by $b = -g \rho/\rho_0$, where $g$ is gravitational acceleration. The density is related to the potential temperature $\theta$ by the relation $\rho = \rho_0 [1+\beta_T (\theta_0 - \theta)]$, where $\beta_T$ is the thermal expansion coefficient and $\theta_0$ is a reference temperature. In Eq.~\eqref{concentration_cont}, ${\bf c}$ denotes the vector of Eulerian concentration fields for each of the tracers. The tracers are passive and thus do not impact the dynamics of ${\bf u}$ or $b$. However, they are non-conserved and ${\bf S}$ in Eq.\ (\ref{concentration_cont}) accounts for sources and sinks due to chemical reactions, as outlined in Section \ref{reactions}. Each of the subgrid-scale (SGS) terms in Eqs.~\eqref{mom_eq}--\eqref{concentration_cont} are fluxes from the SGS model used in the LES.  Note that the form of Eq.~\eqref{mom_eq} is obtained by \citet{suzukifoxkemper16}, although it is mathematically identical to the form in \citet{mcwilliams1997}. 

Langmuir turbulence is created in the simulations by the Stokes drift velocity ${\bf u}_\mathrm{s}$, which appears in Eqs.~\eqref{mom_eq}--\eqref{concentration_cont}. This additional forcing term is expressed in the present LES as
\begin{linenomath*}
\begin{equation}\label{stokes_eq}
{\bf u}_\mathrm{s}(z) = u_\mathrm{s} (z) \left[ \cos(\vartheta_\mathrm{s}) \hat{{\bf x}}+\sin(\vartheta_\mathrm{s}) \hat{{\bf y}}\right]\,,
\end{equation}
\end{linenomath*}
where $u_\mathrm{s}(z)$ is the Stokes drift magnitude vertical profile, which decays faster than exponentially from the surface \citep{donelan1985,webb2011}, and $\vartheta_\mathrm{s}$ is the angle of the Stokes drift velocity in the horizontal (i.e., $x$--$y$) plane. Note that, in the present study, ${\bf u}_\mathrm{s}$ is constant in time and the same at all horizontal locations, and thus depends only on $z$.  Wind, $\vartheta_\mathrm{w}$, and Stokes drift, $\vartheta_\mathrm{s}$, directions are taken to be the same in all simulations (thereby representing wind, as opposed to crossing swell, waves) and both wave-spreading and breaking wave effects are neglected \citep{webb2015}. Prior studies \citep{mcwilliams1997,vanroekel2012,hamlington2014,smith2016} have shown that the inclusion of the Stokes drift velocity in Eqs.~\eqref{mom_eq}--\eqref{concentration_cont} leads to the creation of small-scale, counter-rotating Langmuir cells throughout the domain, with the strongest cells occurring close to the surface.

The numerical code used to perform the simulations is the National Center for Atmospheric Research (NCAR) LES model \citep{moeng1984,mcwilliams1997,sullivan2007}. Horizontal spatial derivatives are calculated pseudo-spectrally, while second- and third-order finite differences are used for vertical derivatives of velocity and tracers, respectively. Third-order Runge--Kutta (RK) time-stepping is used with a constant Courant number. Subgrid-scale viscosity, buoyancy diffusivity, and tracer diffusivity are spatially varying according to the scheme outlined by \citet{sullivan1994}. 

\subsection{Reduced Carbonate Chemistry Model\label{reactions}}
The reduced carbonate chemistry model implemented in the LES is based on the mechanism from \citet{zeebe2001} for carbonate reactions in seawater. This mechanism includes seven species concentrations for $\ce{CO2}$, $\ce{HCO3^-}$, $\ce{CO3^2-}$, $\ce{H^+}$, $\ce{OH^-}$, $\ce{B(OH)3}$, and $\ce{B(OH)4^-}$ (denoted $c_1$--$c_7$; see Table \ref{tracer refs.}), plus $\ce{H2O}$, which is assumed to have a constant concentration. The system of seven reactions describing the mechanism is given as \citep{zeebe2001}
\begin{linenomath*}
\begin{eqnarray*}
\label{eq:r1}\ce{CO2} + \ce{H2O} &\ce{<=>[\alpha_1][\beta_1]}& \ce{HCO3^-} + \ce{H^+} \\
\label{eq:r2}\ce{CO2} + \ce{OH^-} &\ce{<=>[\alpha_2][\beta_2]}& \ce{HCO3^-}  \\
\label{eq:r3}\ce{CO3^2-} + \ce{H^+} &\ce{<=>[\alpha_3][\beta_3]}& \ce{HCO3^-} \\
\label{eq:r4}\ce{HCO3^-} + \ce{OH^-} &\ce{<=>[\alpha_4][\beta_4]}& \ce{CO3^2-} + \ce{H2O} \\
\label{eq:r5}\ce{H2O} &\ce{<=>[\alpha_5][\beta_5]}& \ce{H^+} + \ce{OH^-} \\
\label{eq:r6}\ce{B(OH)3} + \ce{OH^-} &\ce{<=>[\alpha_6][\beta_6]}& \ce{B(OH)4^-} \\
\label{eq:r7}\ce{CO3^2-}+ \ce{B(OH)3} + \ce{H2O} &\ce{<=>[\alpha_7][\beta_7]}& \ce{B(OH)4^-} + \ce{HCO3^-}
\end{eqnarray*}
\end{linenomath*}
where $\alpha_i$ and $\beta_i$ are, respectively, forward and backward reaction coefficients. Table\ \ref{rates_zeebe2001} provides temperature- and salinity-dependent equations for each of these coefficients, as well as values for the coefficients at a temperature of \SI{25}{\degreeCelsius} and salinity of 35 ppt \cite{dickson1994,zeebe2001}. 

\begin{table}[t!]
\centering
\caption{Definition of tracer concentrations $c_i$, terminology, and equilibrium values of tracer concentrations used to initialize the simulations. The equilibrium values correspond to approximate surface values at a temperature of \SI{25}{\degreeCelsius}, salinity of 35 ppt, alkalinity of \SI{2427.89}{\micro\mole\per\kilo\gram}, and DIC concentration of \SI{1992.28}{\micro\mole\per\kilo\gram}.}\label{tracer refs.}
\begin{tabular}{@{}l l l c@{}}
\hline
Tracer  	& Species		& Name & Equilibrium Value (\si{\micro\mole\per\kilo\gram})	\\
\hline
$c_1$ 	& $\ce{CO2}$	& Carbon dioxide & 7.57 				\\
$c_2$ 	& $\ce{HCO3^-}$	& Bicarbonate & \num{1.67e3}	\\
$c_3$	& $\ce{CO3^2-}$	& Carbonate	& \num{3.15e2}\\
$c_4$	& $\ce{H^+}$ & Hydrogen ion & \num{6.31e-3}	\\
$c_5$	& $\ce{OH^-}$ & Hydroxyl & 9.60	\\
$c_6$ 	& $\ce{B(OH)3}$	& Boric acid & \num{2.97e2}	\\
$c_7$ 	& $\ce{B(OH)4^-}$ 	& Tetrahydroxyborate & \num{1.19e2}				\\
\hline
\end{tabular}
\end{table} 

The source terms $S_i$ on the right-hand-side of Eq.~\eqref{concentration_cont} for the rate equation of each tracer $c_i$ are obtained using the law of mass action as
\begin{linenomath*}
\begin{eqnarray}
\hspace{-0.3in}S_1	&=& -(\alpha_1 + \alpha_2 c_5) c_1 + (\beta_1 c_4 + \beta_2) c_2 \,,\label{s1} \\
\hspace{-0.3in}S_2	&=&  (\alpha_1 + \alpha_2 c_5) c_1 - (\beta_1 c_4 + \beta_2 + \beta_3 + \alpha_4 c_5 + \beta_7 c_7 ) c_2  + (\alpha_3 c_4 + \beta_4 +\alpha_7 c_6 ) c_3 \,, \\
\hspace{-0.3in}S_3 	&=& (\beta_3 + \alpha_4 c_5 + \beta_7 c_7 ) c_2 - (\alpha_3 c_4 + \beta_4 + \alpha_7 c_6 ) c_3 \,, \\
\hspace{-0.3in}S_4 	&=& \alpha_1 c_1 - (\beta_1 c_4 - \beta_3)c_2 - \alpha_3 c_4 c_3 + (\alpha_5 - \beta_5 c_4 c_5) \,, \\ 
\hspace{-0.3in}S_5 	&=& -\alpha_2 c_5 c_1+(\beta_2-\alpha_4 c_5) c_2 + \beta_4 c_3 + (\alpha_5 - \beta_5 c_4 c_5)-(\alpha_6 c_5 c_6 - \beta_6 c_7) \,, \\
\hspace{-0.3in}S_6	&=& \beta_7 c_7 c_2 - \alpha_7 c_6 c_3 - (\alpha_6 c_5 c_6 -\beta_6 c_7)\,,	\\
\hspace{-0.3in}S_7	&=& -\beta_7 c_7 c_2 + \alpha_7 c_6 c_3 + (\alpha_6 c_5 c_6 -\beta_6 c_7)\,. \label{s7}	
\end{eqnarray}
\end{linenomath*}
The system of rate equations resulting from this reaction mechanism is, however, numerically stiff and requires a prohibitively small time step to accurately and stably integrate within NCAR LES using the native third-order RK scheme. To overcome this difficulty, two measures were taken: (\emph{i}) a computational singular perturbation analysis and subsequent quasi-steady-state approximation were used to reduce the chemical mechanism, and (\emph{ii}) a Runge--Kutta--Chebyshev scheme was used to integrate the resulting system of rate equations. 

\begin{table}[t!]
\centering
\caption{Temperature and salinity dependent reaction coefficient equations and values at a temperature of \SI{25}{\degreeCelsius} and salinity of 35 ppt for the carbonate chemistry model used in the present study. All values and expressions are taken from \citet{zeebe2001}. Here $A_1 = \SI{4.70e7}{\kilo\gram\per\mole\per\second}$, 
$E_1 = \SI{23.2}{\kilo\joule\per\mole}$, 
$A_6 = \SI{4.58e10}{\kilo\gram\per\mole\per\second}$,
$E_6 = \SI{20.8}{\kilo\joule\per\mole}$, 
$A_7 = \SI{3.05e10}{\kilo\gram\per\mole\per\second}$, and 
$E_7 = \SI{20.8}{\kilo\joule\per\mole}$. 
The temperature and salinity dependent equilibrium constant equations for $K_1^*$, $K_2^*$, $K_W^*$, and $K_B^*$ are given by \citet{dickson1994}. }\label{rates_zeebe2001}
\begin{tabular}{@{}lccc@{}}
\hline
Symbol   				& Equation 										& Value				& Units	\\
\hline
$\alpha_1$ 			& $\exp[1246.98 - \num{6.19e4} / \theta - 183.0 \ln(\theta)]$ 	& 0.037  		& \si{\per\second} \\
$\beta_1$ 			& $\alpha_1/ K_1^*$ 							& \num{2.66e4} 	& \si{\kilo\gram\per\mole\per\second}\\
$\alpha_2$ 			& $A_1 \exp(-E_1/R \theta)$ 						& \num{4.05e3} 	& \si{\kilo\gram\per\mole\per\second}\\
$\beta_2$ 			& $\alpha_2 K_W^*/K_1^*$ 						& \num{1.76e-4} & \si{\per\second}\\
$\alpha_3$ 			& constant 									& \num{5.0e10} 	& \si{\kilo\gram\per\mole\per\second}\\
$\beta_3$ 			& $\alpha_3 K_2^*$ 							& \num{59.4} 	& \si{\per\second}\\
$\alpha_4$ 			& constant 									& \num{6.0e9} 	& \si{\kilo\gram\per\mole\per\second}\\
$\beta_4$ 			& $\alpha_4 K_W^* / K_2^*$ 					& \num{3.06e5} 	& \si{\per\second}\\
$\alpha_5$ 			& constant 									& \num{1.40e-3} & \si{\kilo\gram\per\mole\per\second}\\
$\beta_5$ 			& $\alpha_5/ K_W^*$ 							& \num{2.31e10} & \si{\kilo\gram\per\mole\per\second}\\
$\alpha_6$ 			& $A_6 \exp(-E_7/R \theta)$ 						& \num{1.04e7} 	& \si{\kilo\gram\per\mole\per\second}\\
$\beta_6$ 			& $\alpha_6 K_W^* / K_B^*$ 					& \num{249} 	& \si{\per\second}\\
$\alpha_7$ 			& $A_7 \exp(-E_8/R \theta)$ 						& \num{6.92e6} 	& \si{\kilo\gram\per\mole\per\second}\\
$\beta_7$ 			& $\alpha_7 K_2^* / K_B^*$ 					& \num{3.26e6} 	& \si{\kilo\gram\per\mole\per\second}\\
\hline
\end{tabular}
\label{default}
\end{table}

\subsubsection{Computational Singular Perturbation Analysis}
Computational singular perturbation (CSP) analysis \citep{lam1988,lam1994,goussis1992,lam1993} was applied to the chemical kinetic system represented by Eqs.~\eqref{s1}--\eqref{s7} to identify candidate species for quasi-steady state (QSS) approximations following the approach outlined by \citet{lu2008a,lu2008b} and \citet{niemeyer2015}. To perform the CSP analysis, the reaction rate equations from Eqs.~\eqref{concentration_cont} and \eqref{s1}--\eqref{s7} were first written as a zero-dimensional (i.e., only time dependent) system given by
\begin{linenomath*}
\begin{equation}
\frac{d\mathbf{c}}{dt} 	= \mathbf{S}(\mathbf{c})\quad \Rightarrow \quad \frac{d\mathbf{S}}{dt}= \mathbf{J} \mathbf{S}\,, 
\end{equation}
\end{linenomath*}
where $\mathbf{J} = \partial \mathbf{S}/\partial \mathbf{c}$ is the Jacobian matrix. The CSP analysis decomposes the source terms $\mathbf{S}$ into a vector of modes $\mathbf{f}$ using row basis vectors $\mathbf{B}$ as $\mathbf{f} = \mathbf{B} \mathbf{S}$. The time derivative of $\mathbf{f}$ then gives
\begin{linenomath*}
\begin{equation}\label{feq}
\frac{d \mathbf{f}}{d t} = \left(\frac{d\mathbf{B}}{dt}+\mathbf{B}\cdot\mathbf{J}\right)\mathbf{S} = \mathbf{\Lambda} \mathbf{f}\,,
\end{equation}
\end{linenomath*}
where $\mathbf{\Lambda}$ is given by
\begin{linenomath*}
\begin{equation}
 \mathbf{\Lambda}=\left(\frac{d\mathbf{B}}{dt}+\mathbf{B}\mathbf{J}\right)\mathbf{A}\,,
\end{equation}
\end{linenomath*}
and $\mathbf{A}=\mathbf{B}^{-1}$. For simplicity, the Jacobian matrix was assumed to be time independent such that $d\mathbf{B}/dt = 0$, leading to
\begin{linenomath*}
\begin{equation}\label{eigen}
\mathbf{\Lambda} = \mathbf{B} \mathbf{J} \mathbf{A}\,,
\end{equation}
\end{linenomath*}
where $\mathbf{\Lambda}$ contains the eigenvalues of $\mathbf{J}$ on the diagonal. The eigendecomposition of the Jacobian was performed using the NumPy function \texttt{numpy.linalg.eig}~\citep{vanderwalt2011}; the Jacobian itself was evaluated analytically using SymPy~\citep{Meurer2017}. The CSP basis vectors $\mathbf{A}$ and $\mathbf{B}$ are then the right and left eigenvectors of $\mathbf{J}$, respectively.

Using the eignevalues from Eq.~\eqref{eigen}, the system dynamics were separated into fast and slow subspaces, where the evolution of the modes $\mathbf{f}$ in each subspace is given from Eq.~\eqref{feq} by \cite{niemeyer2015}
\begin{linenomath*}
\begin{equation}
\frac{d}{dt}\left[\begin{array}{c} \mathbf{f}^\mathrm{fast} \\ \mathbf{f}^\mathrm{slow}\end{array}\right] = \left[\begin{array}{cc} \mathbf{\Lambda}^\mathrm{fast} & {} \\ {} & \mathbf{\Lambda}^\mathrm{slow}\end{array}\right]\left[\begin{array}{c} \mathbf{f}^\mathrm{fast} \\ \mathbf{f}^\mathrm{slow}\end{array}\right]\,.
\end{equation}
\end{linenomath*}
The fast modes $\mathbf{f}^\mathrm{fast}$ decay rapidly and have negative eigenvalues $\mathbf{\Lambda}^\mathrm{fast}$ that are much larger in magnitude than the eigenvalues $\mathbf{\Lambda}^\mathrm{slow}$ associated with the slow subspace $\mathbf{f}^\mathrm{slow}$. 

Formally, the fast and slow subspaces were identified by defining a cutoff time scale $\tau_\text{c} / \gamma_{\text{CSP}}$, where $\tau_\text{c}$ is a characteristic time scale of the global system dynamics and $\gamma_{\text{CSP}}$ is a safety factor. The two subspaces were then separated by requiring that the time scale associated with the smallest magnitude eigenvalue in the fast subspace (corresponding to the slowest mode in the fast subspace), denoted $\lambda_\mathrm{min}(\mathbf{\Lambda}^\mathrm{fast})$, be less than the cutoff time scale:
\begin{linenomath*}
\begin{equation}
\frac{-1}{\lambda_{\min} \left(\mathbf{\Lambda}^{\text{fast}} \right)} <\frac{\tau_c}{\gamma_{\text{CSP}}}\,.
\label{eq:csp_separation}
\end{equation}
\end{linenomath*}
The negative-valued eigenvalues in $\mathbf{\Lambda}$ with magnitudes greater than $\lambda_\mathrm{min}$ are all part of the fast subspace, while the remaining eigenvalues are part of the slow subspace. From $\mathbf{\Lambda}^\mathrm{fast}$ and $\mathbf{\Lambda}^\mathrm{slow}$, it was then possible to identify the fast and slow subspace modes, $\mathbf{f}^\mathrm{fast}$ and $\mathbf{f}^\mathrm{slow}$, respectively. The characteristic time $\tau_\text{c}$ was defined to be the relaxation time for \ce{CO2} to reach 1\% of its equilibrium concentration after a \SI{1}{\micro\mole\per\kilo\gram} increase in \ce{CO2}, a \SI{1}{\micro\mole\per\kilo\gram} decrease in \ce{CO^2-_3}, and a \SI{2}{\micro\mole\per\kilo\gram} increase in \ce{OH-}~\cite{zeebe2001}, giving $\tau_\text{c} = \SI{63.03}{\second}$. The safety factor was set as $\gamma_{\text{CSP}} = 50$.  

\subsubsection{Quasi-Steady-State Approximation}
Projecting $\mathbf{S}$ onto the fast and slow subspaces gives $\mathbf{S} = \mathbf{S}^\mathrm{fast}+\mathbf{S}^\mathrm{slow}$, where $\mathbf{S}^\mathrm{fast}=\mathbf{Q}^{\mathrm{fast}}\mathbf{S}$ and $\mathbf{S}^\mathrm{slow}=\mathbf{Q}^{\mathrm{slow}}\mathbf{S}$. Here, $\mathbf{Q}^{\mathrm{fast}}$ and $\mathbf{Q}^{\mathrm{slow}}$ are, respectively, the fast and slow projection matrices given by
\begin{linenomath*}
\begin{equation}
\mathbf{Q}^{\mathrm{fast}} = \mathbf{A}^{\text{fast}} \mathbf{B}^{\text{fast}} \,, \quad \mathbf{Q}^{\mathrm{slow}} = \mathbf{A}^{\text{slow}} \mathbf{B}^{\text{slow}}\,.
\end{equation}
\end{linenomath*}
The basis vectors were split into fast- and slow-mode vectors by applying Eq.~\eqref{eq:csp_separation} to identify the associated fast and slow eigenvalues, and $\mathbf{Q}^{\mathrm{slow}}$ was constructed. Then, species were identified as good candidates for the QSS assumption if they correlated (or projected) weakly to the slow subspace. For the $i$th species, this was determined using
\begin{linenomath*}
\begin{equation}
\left|  \mathbf{Q}_{i,i}^{\text{slow}} \right| < \epsilon_{\text{CSP}} \,,
\label{eq:csp_criterion}
\end{equation}
\end{linenomath*}
where $\mathbf{Q}_{i,i}^{\text{slow}}$ is the $i$th diagonal element of $\mathbf{Q}^{\text{slow}}$ and $\epsilon_{\text{CSP}}$ is a small threshold value (0.1 was used here). Practically, species were determined to satisfy the criterion given by Eq.~\eqref{eq:csp_criterion} by calculating the maximum values of $\mathbf{Q}_{i,i}^{\text{slow}}$ for all species over a simulated relaxation back to equilibrium after a 10\% perturbation to the concentration of \ce{CO2}, at a temperature of \SI{25}{\celsius} and salinity of 35 ppt.

\begin{figure}[t!]
\centering
\includegraphics[width=0.45\linewidth]{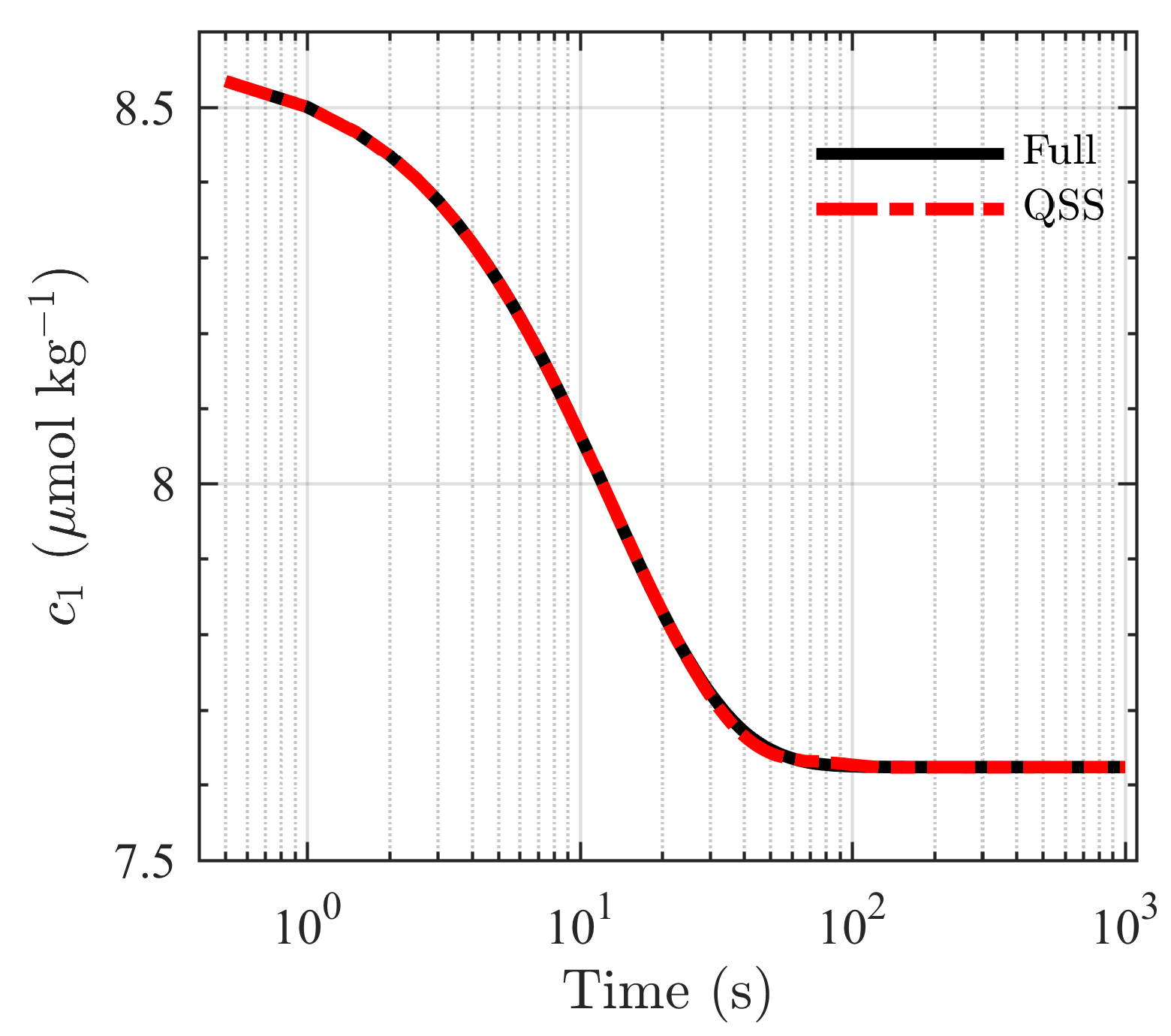}
\caption{Relaxation times for the concentration of \ce{CO2}, denoted $c_1$, determined using the full kinetic model (solid black line) and using the reduced model with the QSS approximation in Eq.~\eqref{qss} applied to \ce{H^+} (dash-dot red line).}
\label{fig:relaxation_compare}
\end{figure}

The CSP analysis identified two QSS candidates: \ce{H^+} and \ce{OH^-}, with slow-subspace contributions (i.e., $\mathbf{Q}_{i,i}^{\text{slow}}$) of \num{1.81e-5} and \num{2.72e-2} respectively. 
Both species satisfied the criterion in Eq.~\eqref{eq:csp_criterion}, but it was found that the approximation could only be applied to \ce{H^+} without introducing significant error. Thus, the CSP analysis determined that \ce{H^+} (which is connected to the pH) was a candidate for the QSS approximation by identifying it as a ``radical'' (in the CSP context) because it contributed little to the slow, controlling modes of the system dynamics, below a safety factor. 

Using the QSS approximation for the concentration of \ce{H^+} (tracer $c_4$, see Table \ref{tracer refs.}), it was assumed that $S_4=0$ and that $c_4$ could be obtained algebraically as
\begin{linenomath*}
\begin{equation}\label{qss}
c_4^* = \frac{\alpha_1 c_1 + \beta_3 c_2 + \alpha_5}{\beta_1 c_2 + \alpha_3 c_3 + \beta_5 c_5}\,,
\end{equation}
\end{linenomath*}
where $c_4^*$ denotes the QSS approximation for $c_4$. Computationally, the resulting reduced chemical mechanism was less stiff due to the use of QSS for one of the three fastest-evolving species, allowing a 50\% increase in the time step required for the simulations, and required the integration of only six, as opposed to seven, coupled differential equations [i.e., no differential equation needed to be integrated for $c_4$, since this tracer concentration was given algebraically by Eq.~\eqref{qss}]. 

The error due to the QSS assumption was estimated using a zero-dimensional test where the system was perturbed by an increase of \SI{1}{\micro\mole\per\kilo\gram} increase in $c_1$, a \SI{1}{\micro\mole\per\kilo\gram} decrease in $c_3$ (to maintain constant DIC concentration), and a \SI{2}{\micro\mole\per\kilo\gram} increase in $c_5$ (to maintain constant alkalinity)~\cite{zeebe2001}, after which all species relaxed back to their respective equilibrium values. Examining the temporal evolution, the concentrations of all species in the reduced model agreed within \num{1e-5}\% of the full model [except for the concentration of $c_4$, which was analytically provided by Eq.\ (\ref{qss}) resulting from the QSS assumption] over the entire equilibration period ($\sim$\SI{60}{\second}). Figure~\ref{fig:relaxation_compare} shows the results from this test.

\subsubsection{Runge--Kutta--Chebyshev Solver}
In the simulations, time integration of the advection and chemistry was split \cite{strang1968} such that the advection remained within the pre-existing third-order RK scheme in NCAR LES and the chemistry was integrated in two half steps, before and after the advection step. The chemistry integration used an explicit second-order Runge--Kutta--Chebyshev (RKC) scheme that is robust for moderately stiff equations \cite{sommeijer1997,verwer2004,niemeyer2014}. While explicit, the RKC algorithm is stabilized to handle more stiffness than traditional RK methods. The RKC scheme is explicit and constructed like other multistage explicit RK methods, but uses an increased, variable number of stages and coefficients chosen to increase the stability region rather than accuracy---thus the method is known as a stabilized explicit scheme. The use of the RKC solver provided an additional increase in the time step required for the simulations, from roughly $10^{-5}$ s without the RKC  solver (i.e., using the native third-order RK scheme in NCAR LES) to roughly $0.1$ s with the RKC solver.

\subsection{Physical Setup \label{phys scenario}}
The physical and computational parameters used to setup the simulations are summarized in Table \ref{tab:setup}. All simulations were initialized with a mixed layer depth of \SI{30}{\meter}, with uniform stratification (i.e., linearly increasing density) below. Buoyancy, density, and temperature were all spatially and temporally varying in the simulations, but salinity was assumed fixed at 35 ppt. The physical domain size was $L_x\times L_y \times L_z = 320\times 320\times \SI{96}{\meter\cubed}$ with a horizontal ($x-y$) resolution of \SI{2.5}{\meter} and a vertical ($z$) resolution of \SI{0.75}{\meter}. The initial velocities were motionless. Periodic boundary conditions were used in horizontal directions and a zero vertical velocity condition was applied at the bottom boundary. A surface wind stress of \SI{0.025}{\newton\per\square\meter} was applied to all simulations along the $x$ direction, with a friction velocity of $u_\tau = \SI{5.3e-3}{\meter\per\second}$, corresponding to a \SI{10}{\meter} wind speed of \SI{5.75}{\meter\per\second}. Zero-gradient boundary conditions were used for the temperature at the top and bottom of the domain, and the diurnal cycle was not modeled in the simulations.  

\begin{table}[b!]
\caption{Summary of physical and computational parameters used in the numerical simulations.}\label{tab:setup}
\centering
\begin{tabular}{@{}l c c c c@{}}
\hline
Physical size, $L_x \times L_y \times L_z$ 					& \multicolumn{4}{c}{\SI{320}{\meter}$\times$\SI{320}{\meter}$\times$\SI{-96}{\meter}} \\
Grid size, $N_x \times N_y \times N_z$						& \multicolumn{4}{c}{128$\times$128$\times$128}	\\
Grid Resolution, $\triangle_x \times \triangle_y \times \triangle_z$	& \multicolumn{4}{c}{\SI{2.5}{\meter}$\times$\SI{2.5}{\meter}$\times$\SI{0.75}{\meter}}	\\
Reference density, $\rho_0$ 								& \multicolumn{4}{c}{\SI{1000}{\kilo\gram\per\cubic\meter}} \\
Thermal expansion coefficient, $\beta_T$ 					& \multicolumn{4}{c}{\SI{2e-4}{\per\kelvin}} \\
Coriolis parameter, $\textbf{f}_\mathrm{c}$ 					& \multicolumn{4}{c}{\SI{0.729e-4}{\per\second}$\hat{\textbf{z}}$} \\
Initial mixed layer depth, $H_{\text{ML},0}$						& \multicolumn{4}{c}{\SI{-30}{\meter}} \\
Wind speed at 10 m, $U_{10}$ 								& \multicolumn{4}{c}{\SI{5.75}{\meter\per\second}} \\
Stokes drift direction, $\vartheta_\mathrm{s}$          				& \multicolumn{4}{c}{\SI{0}{\degree}} \\
Water-side wind friction velocity, $u_\tau=\sqrt{\tau/\rho_o}$	& \multicolumn{4}{c}{\SI{5.3e-3}{\meter\per\second}} \\
Wind stress, $\tau$ 										& \multicolumn{4}{c}{\SI{0.025}{\newton\per\square\meter}} \\
Wind direction, $\vartheta_\mathrm{w}$ 								& \multicolumn{4}{c}{\SI{0}{\degree}} \\
\hline
Surface stokes drift, $u_\mathrm{s}(0)$ (\si{\meter\per\second}) 	& 0.000  	& 0.032 	& 0.080 	& 0.132 \\
Langmuir number, $\mathrm{La}_\mathrm{t}  \equiv [u_\tau/u_\mathrm{s}(0)]^{1/2}$	& $\infty$	& 0.40	& 0.30	& 0.20 \\
Simulation label 									   		& NS		& La04	& La03	& La02 \\
\hline
\end{tabular}
\end{table}

\begin{figure}[t!]
\centering
\includegraphics[width=0.45\linewidth]{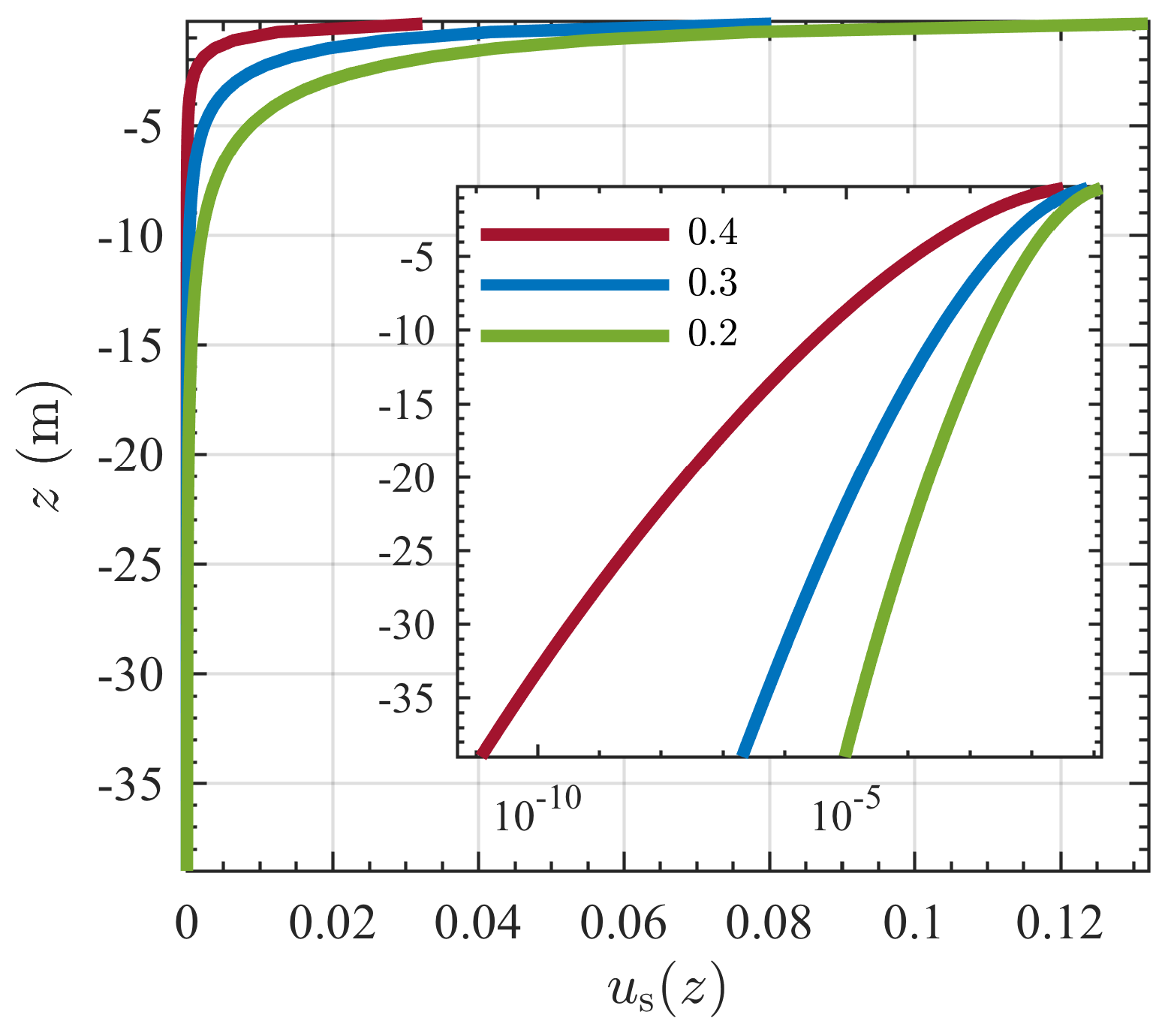}
\caption{Stokes drift velocity $u_\mathrm{s}(z)$ as a function of depth $z$ for Langmuir numbers $\text{La}_\mathrm{t} = 0.4$, 0.3, and 0.2 (red, blue, and green lines, respectively), where $\mathrm{La}_\mathrm{t}  \equiv [u_\tau/u_\mathrm{s}(0)]^{1/2}$). The main plot shows $u_\mathrm{s}(z)$ on linear axes and the inset shows $u_\mathrm{s}(z)$ on semilog axes.}
\label{figure1}
\end{figure}

Four wave-forcing scenarios were examined by varying the Stokes drift velocity: a single case with no Langmuir turbulence and three cases with increasing strengths of Langmuir turbulence. The Stokes drift velocity profiles, $u_\mathrm{s}(z)$, applied in each of the Langmuir cases are shown in Figure~\ref{figure1}. The strength of the Langmuir turbulence is characterized by its turbulent Langmuir number, $\text{La}_\mathrm{t}^2 = u_\tau / u_\mathrm{s}(0)$, where $u_\mathrm{s}(0)$ is the surface Stokes drift from each of the profiles shown in Figure~\ref{figure1}. The four scenarios examined correspond to $\mathrm{La}_\mathrm{t} = \infty$, 0.4, 0.3, and 0.2, where $\mathrm{La}_\mathrm{t} = \infty$ is the non-Langmuir case. The range of Langmuir numbers explored here is realistic \citep{li2016,li2017}, and 0.3 is the value attained under fully developed seas \citep{webb2011}. The $\mathrm{La}_\mathrm{t} = 0.4$ and $0.2$ cases are intended to reveal the effects of weaker and stronger Langmuir turbulence, respectively, as compared to the baseline value. Additional discussion of the physical setup represented by these simulations, as well as how turbulence-chemistry interactions would vary for different conditions, is provided in Section \ref{time}.
    
According to \citet{callaghan2008}, less than 0.25\% of the global sea surface area is expected to be covered by whitecapping (i.e., breaking waves) for the wind strength considered here. Consequently, no wave-breaking parameterization was used in these simulations, although one has been developed for the NCAR LES model \citep{sullivan2007} and could be explored in future work. Similarly, bubble parameterizations were not included in these simulations \citep{liang2011}, although the effects of bubbles are likely to be significant \citep{woolf1993}, particularly given their connection to Langmuir turbulence \citep{farmer1995,thorpe2003}. Although the present study is specifically focused on the effects of enhanced vertical mixing by Langmuir turbulence, future work will explore the effects of bubbles, for example using the parameterization for bubble-enhanced air-sea fluxes given by \citet{woolf1993}.

For each physical scenario, species concentrations were initialized uniformly throughout the domain using equilibrium values for a temperature of \SI{25}{\degreeCelsius}, salinity of 35 ppt, alkalinity of \SI{2427.89}{\micro\mole\per\kilo\gram}, and DIC of \SI{1992.28}{\micro\mole\per\kilo\gram} (see Table~\ref{tracer refs.}). Here, DIC is the sum of all carbon containing species and its concentration, denoted $c_\mathrm{DIC}$, is defined as $c_\mathrm{DIC}\equiv c_1+c_2+c_3$. Each tracer was subject to periodic boundaries in horizontal directions with, initially, no vertical fluxes at the bottom and top boundaries.

\subsection{Simulation Procedure}    
After approximately seven days during which turbulence was allowed to develop and tracers relaxed to their equilibrium values based on the local value of the temperature (salinity is fixed), additional \ce{CO2} was allowed to enter through the top boundary according to Henry's law for gas flux across the air-sea interface \citep{wanninkhof1992}. This flux law is given as \citep{smith2016}
\begin{linenomath*}
\begin{equation}\label{flux}
F_{\ce{CO2}}(x,y,t) = k_{\ce{CO2}} \left[c_{1}^{\text{air}} -  c_1(x,y,0,t)\right]\,,
\end{equation}
\end{linenomath*}
where $F_{\ce{CO2}}$ is the downward flux rate across the boundary, which varies over horizontal directions and time as temperature and $c_1$ vary, $k_{\ce{CO2}}$ is the species flux rate (or piston velocity), $c_{1}^{\text{air}}$ is the concentration in air, and $c_1(x,y,0,t)$ is the concentration just below the surface. The value of $c_{1}^{\text{air}}$ was fixed at a 10\% increase above the initial mixed layer average of $c_1$ (namely, $c_{1}^{\text{air}}$ = \SI{8.3}{\micro\mole\per\kilo\gram}, see Table~\ref{tracer refs.}). The piston velocity, $k_{\ce{CO2}}$, is given as a function of the 10 m wind speed $U_{10}$ (see Table \ref{tab:setup}) and Schmidt number $\mathrm{Sc}$ as
\begin{linenomath*}
\begin{equation}\label{henry}
k_{\ce{CO2}} = 0.31 U_{10}^2 \sqrt{\frac{660}{\mathrm{Sc}}}\,,
\end{equation}
\end{linenomath*}
where $\mathrm{Sc}$ is a function of temperature given by \citep{wanninkhof1992}
\begin{linenomath*}
\begin{equation}\label{schmidt}
\mathrm{Sc}= 2073.1-125.62\theta + 3.6276\theta^2-0.043219\theta^3\,.
\end{equation}
\end{linenomath*}
Note that in the above expressions, $k_{\ce{CO2}}$ has units of \SI{}{\centi\meter\per\hour} and $\theta$ in Eq.~\eqref{schmidt} is assumed to have units of \SI{}{\degreeCelsius}. The piston velocity $k_{\ce{CO2}}$ from Eq.~\eqref{henry} does not include a bubble parameterization, but consideration of bubbles, as well as their coupling to Langmuir turbulence, is an important direction for future research. 
     
The simulations were run for six additional hours after initiating the air-sea flux of \ce{CO2} and analysis of the data was carried out after this period. Longer simulations were not performed due to the computational expense of integrating the time-dependent chemistry, and also due to the artificiality of neglecting the diurnal cycle over long periods. Six hours was found to be sufficient for identifying trends in the data, but all of the conclusions contained herein should be understood as only strictly valid up to six hours; future work is necessary to determine carbonate chemistry evolution over much longer time periods, including diurnal and seasonal cycles.
     
Two additional sets of simulations were also performed: one in which each of the chemical species concentrations were calculated at carbonate chemical equilibrium \citep{zeebe2001} and one in which there were no chemical reactions, but still including surface fluxes, transport, and mixing. The equilibrium model is implemented by ensuring that, at each location and time, there is no propensity for the concentrations $c_i$ to change due to reactions. This is accomplished by setting $\mathbf{S}$ to zero and solving the system of nonlinear coupled algebraic equations represented by Eqs.~\eqref{s1}--\eqref{s7} to find the equilibrium values of $\mathbf{c}$. The three chemistry models are referred to in the following as the Time-dependent Chemistry (TC), the Equilibrium Chemistry (EC), and the No Chemistry (NC) models, respectively. The physical scenarios for these three sets of simulations were identical and are described in Section \ref{phys scenario}.

The EC model is representative of how carbonate chemistry is most commonly calculated within ESMs, where reactions are assumed to be instantaneous (i.e., infinitely fast) with respect to modeled physical processes. The NC model, by contrast, effectively represents reactions that are infinitely slow. These two sets of simulations thus give upper and lower bounds for ocean carbonate chemistry reaction times.

\section{Results\label{results}} 

\subsection{Physical Ocean State\label{physics}}
\begin{figure}[t!]
\centering
\includegraphics[width=\linewidth]{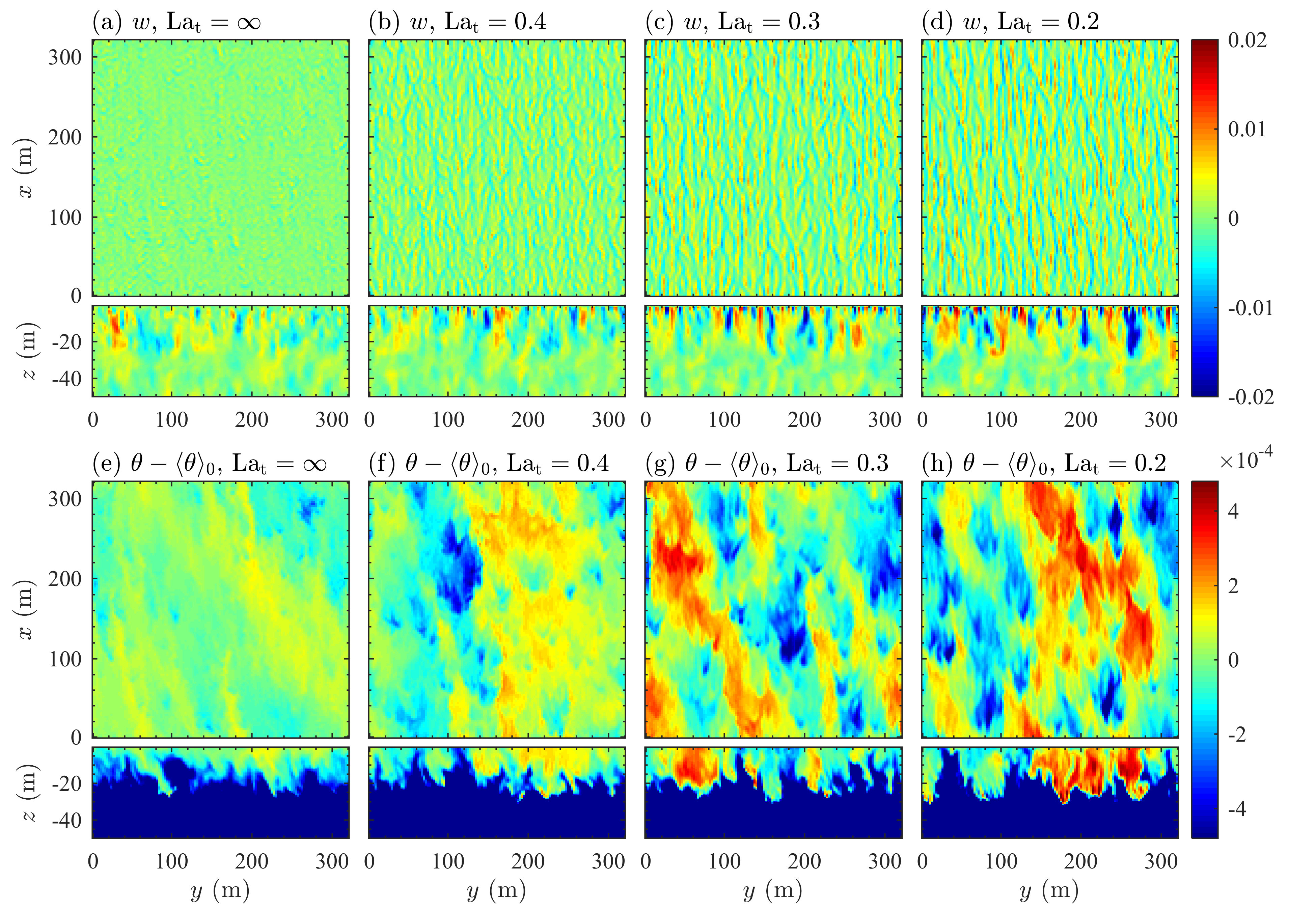}
\caption{Fields of (a-d) vertical velocity $w$ in units of m s$^{-1}$ and (e-h) potential temperature fluctuation $\theta-\langle \theta\rangle_0$ in units of K in horizontal planes at the surface (top subpanels) and in vertical planes in the middle of the domain (bottom subpanels) for Langmuir numbers $\mathrm{La}_\mathrm{t} = \infty$ (a,e), 0.4 (b,f), 0.3 (c,g), and 0.2 (d,h) using time-dependent chemistry. The background potential temperature $\langle \theta\rangle_0$ is computed as the $x$--$y$ average of $\theta$ at the surface.}
\label{fig:figure3}
\end{figure}

Figure~\ref{fig:figure3} shows fields of vertical velocity and potential temperature fluctuations for the non-Langmuir ($\text{La}_\mathrm{t} = \infty$) and three Langmuir ($\text{La}_\mathrm{t} =$ 0.4, 0.3, 0.2) cases. The $x$--$y$ surface fields of vertical velocity in Figures~\ref{fig:figure3}(b-d) for the three Langmuir cases show the streak-like patterns formed by long counter-rotating Langmuir cells that are characteristic of Langmuir turbulence. Although these streaks are spatially variable in direction and magnitude, they are preferentially aligned with the wind direction along the $x$-axis (since $\vartheta_\mathrm{s}=\vartheta_\mathrm{w}=0^\circ$; see Table \ref{tab:setup}) and generally increase in magnitude as $\mathrm{La}_\mathrm{t}$ decreases (i.e., with increasing strength of Langmuir turbulence).

\begin{figure}[t!]
\centering
\includegraphics[width=0.8\linewidth]{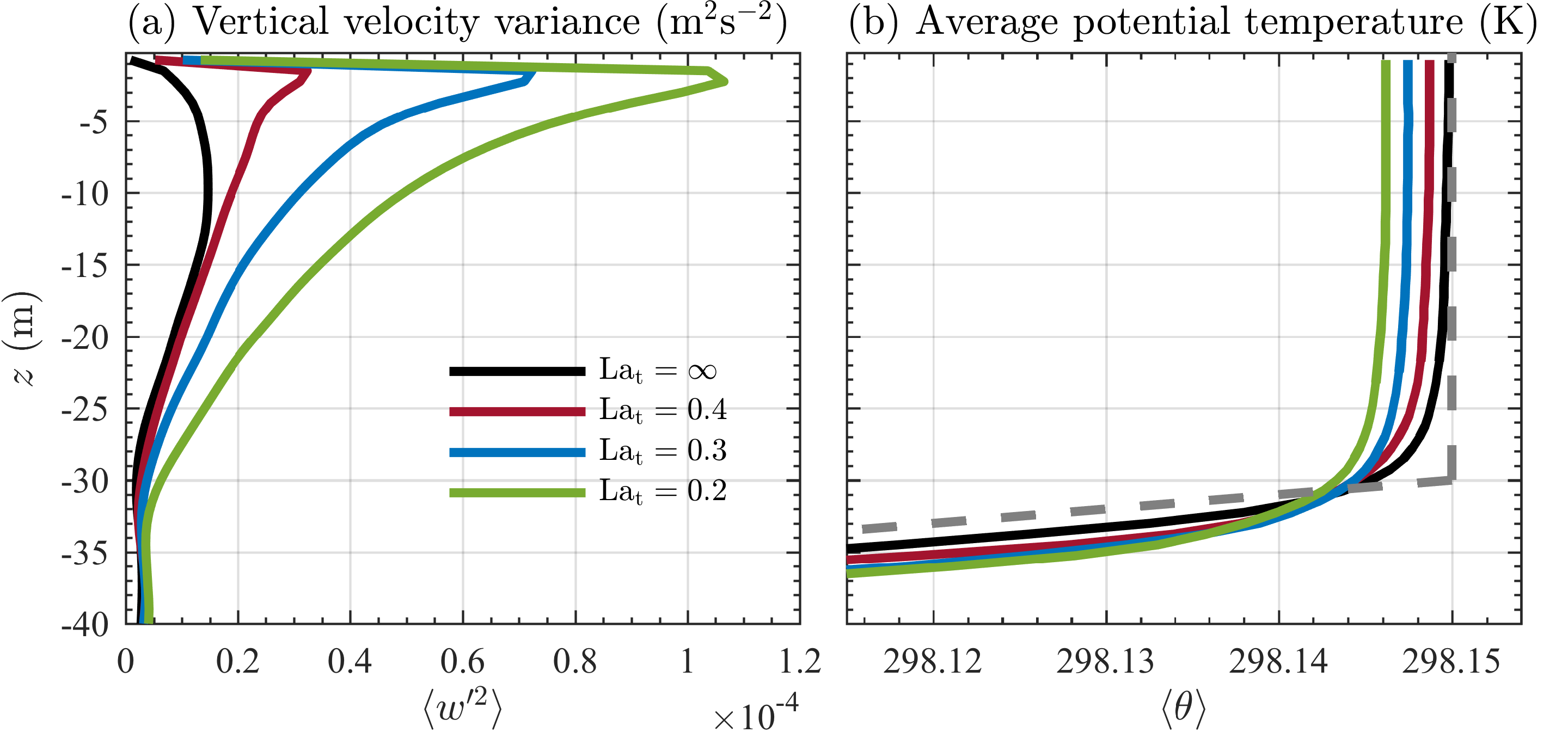}
\caption{Vertical profiles of (a) vertical velocity variance $\langle w'^2\rangle$ and (b) average potential temperature $\langle \theta\rangle$ for Langmuir numbers $\mathrm{La}_\mathrm{t} = \infty$, 0.4, 0.3, and 0.2 (black, red, blue, and green lines, respectively). Statistics are computed in horizontal $x$--$y$ planes as a function of depth $z$. The dashed gray line in (b) shows the initial temperature profile with uniform temperature above $z = \SI{-30}{\meter}$ and constant stratification below.}
\label{fig:figure4}
\end{figure}

The vertical velocities are smallest in the shear-only non-Langmuir case [Figure~\ref{fig:figure3}(a)] and largest in the $\mathrm{La}_\mathrm{t} = 0.2$ case [Figure~\ref{fig:figure3}(d)]. These enhanced vertical velocities are also evident in the $x$--$y$ averaged depth profiles of vertical velocity variance shown in Figure~\ref{fig:figure4}(a), where the peak magnitude of the vertical velocity variance is greatest for the smallest Langmuir number (i.e., $\mathrm{La}_\mathrm{t} = 0.2$), with a progressive increase in magnitude from the non-Langmuir (i.e., $\mathrm{La}_\mathrm{t} = \infty$) case.

Figure~\ref{fig:figure3} further shows that, in addition to the increase in magnitude of vertical mixing, the vertical extent of mixing is greater for the Langmuir cases than for the non-Langmuir case. This enhanced mixing, which increases in strength as Langmuir number decreases, results in a deeper mixed layer. This is indicated by the fluctuating potential temperature fields in Figures~\ref{fig:figure3}(e-h), which show that, as the Langmuir number decreases, greater temperature fluctuations are observed throughout the mixed layer, and the mixed layer extends to slightly greater depths. 

The deepening of the mixed layer is perhaps more evident in the $x$--$y$ averaged potential temperature profiles shown in Figure~\ref{fig:figure4}(b). All simulations begin with the same temperature profile (the dashed gray line in Figure~\ref{fig:figure4}) and all deviate from this initial profile by the end of the seven day spin-up period, but the deviation becomes increasingly pronounced as the strength of Langmuir turbulence increases. In particular, the increased mixing associated with Langmuir turbulence has deepened the mixed layer by approximately \SIrange{1}{3}{\meter}, depending on the case, over the course of the spin-up. 

This deepening not only increases the total volume of the mixed layer, thereby increasing the short-term new carbon reservoir size, but also decreases the average temperature of the mixed layer by entraining cooler waters from below. While this decrease in temperature may not seem substantial in the larger context, carbonate chemistry and air-sea gas fluxes are both sensitive to temperature, as indicated by the temperature-dependent reaction rate coefficients in Table \ref{rates_zeebe2001} and the Henry's law gas flux expression in Eq.~(\ref{henry}). As temperatures cool, \ce{CO2} becomes more soluble in water, allowing more \ce{CO2} to enter the domain. However, reaction times also decrease, leaving carbon as \ce{CO2} longer before it is converted into $\ce{HCO3^-}$ and $\ce{CO3^2-}$. This effect of temperature (i.e., the competition between increased solubility and decreased reaction times), is not directly examined in this study, although future studies exploring these effects within the context of Langmuir turbulence and carbonate chemistry are certainly warranted. 

\subsection{Effects of Langmuir Turbulence on Carbonate Chemistry\label{langmuir effect}}
\begin{figure}[t!]
\centering
\includegraphics[width=\linewidth]{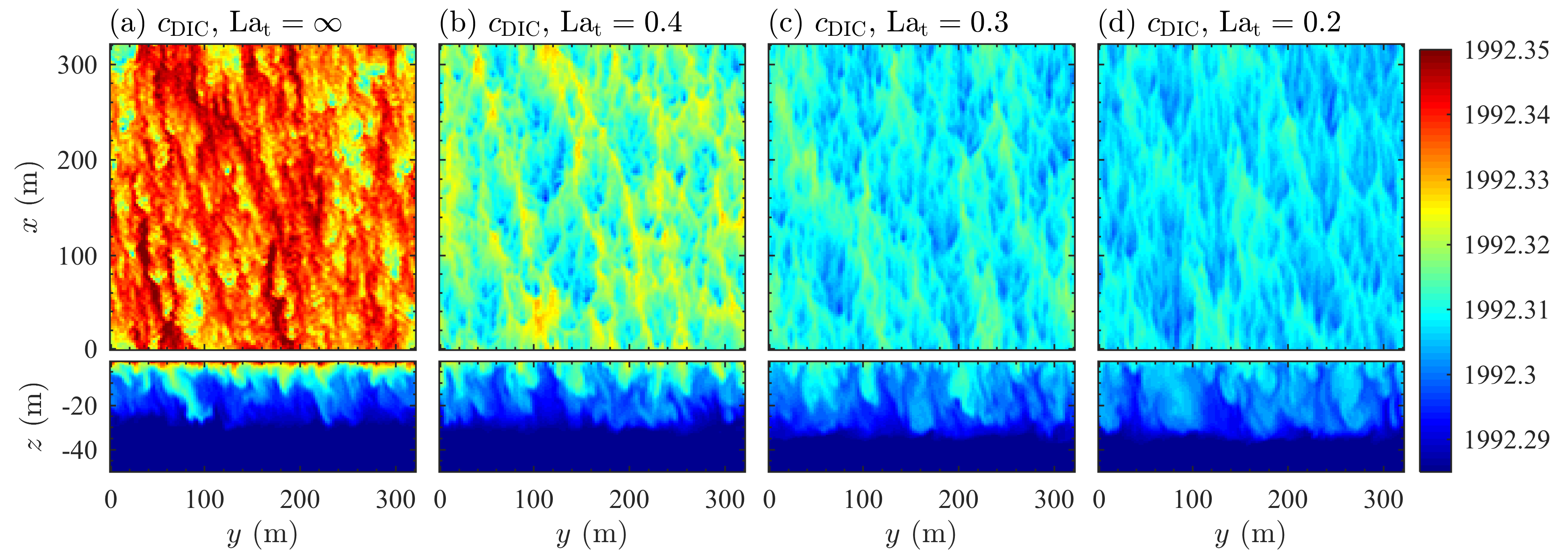}
\caption{Fields of DIC concentration $c_\mathrm{DIC}$ in units of $\mu$mol kg$^{-1}$ in horizontal planes at the surface (top subpanels) and in vertical planes in the middle of the domain (bottom subpanels) for Langmuir numbers $\mathrm{La}_\mathrm{t} = \infty$, 0.4, 0.3, and 0.2 (a-d) using time-dependent chemistry.}
\label{fig:figure5}
\end{figure}

As Langmuir turbulence strengthens, additional carbon is brought through the surface and progressively further down into the mixed layer, as shown in Figure~\ref{fig:figure5}. In particular, Figure~\ref{fig:figure5} shows that the vertical extent of $c_\mathrm{DIC}$ distribution in the mixed layer increases as $\mathrm{La}_\mathrm{t}$ decreases, while the peak surface concentrations decrease. This is partially due to the fast mixing and increased vertical flux associated with Langmuir turbulence, but also to the mixed layer deepening effect of Langmuir turbulence \citep{hamlington2014}; both of these physical effects were described in Section \ref{physics}.

\begin{figure}[t!]
\centering
\includegraphics[width=\linewidth]{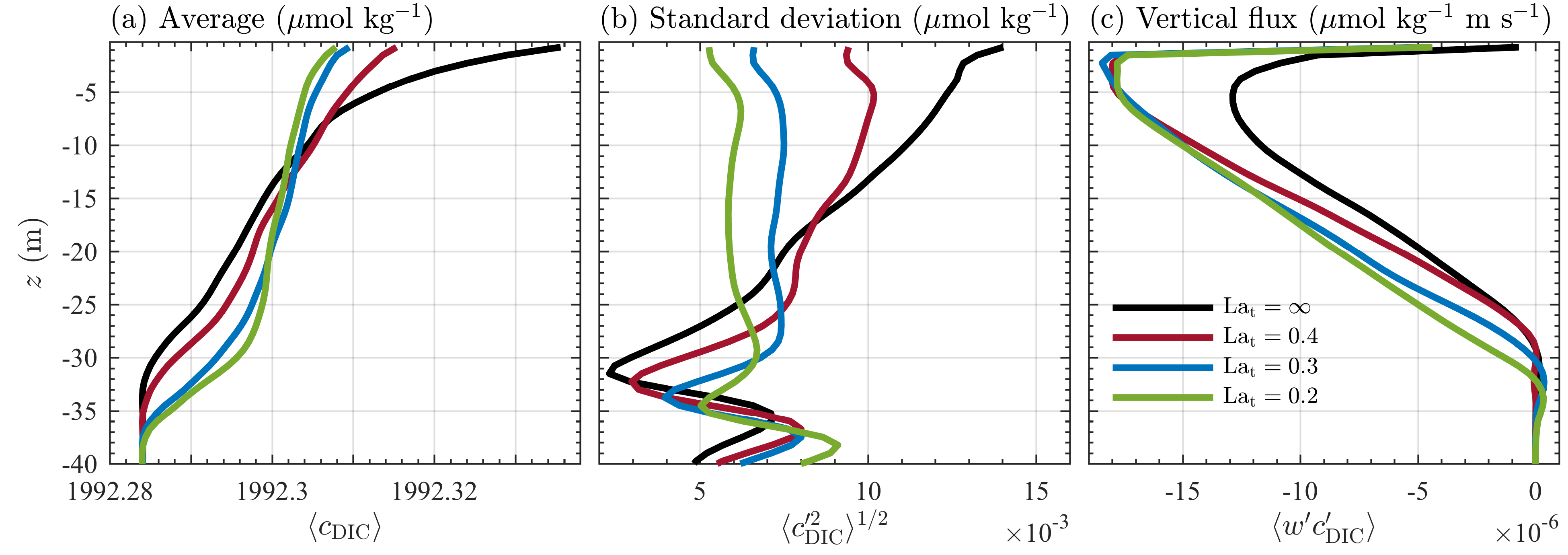}
\caption{Vertical profiles of (a) average $c_\mathrm{DIC}$, (b) standard deviation of $c_\mathrm{DIC}$, and (c) vertical flux of $c_\mathrm{DIC}$ for Langmuir numbers $\mathrm{La}_\mathrm{t} = \infty$, 0.4, 0.3, and 0.2 (black, red, blue, and green lines, respectively), using time-dependent chemistry. Statistics are computed in horizontal $x$--$y$ planes as a function of depth $z$.}
\label{fig:figure6}
\end{figure}

Figures~\ref{fig:figure6}(a) and (b) show the $x$--$y$ average and standard deviation, respectively, of $c_\mathrm{DIC}$ as a function of depth for $\text{La}_\mathrm{t} = \infty$, 0.4, 0.3, and 0.2. In the non-Langmuir case (i.e., $\text{La}_\mathrm{t} = \infty$), there is a much greater concentration and standard deviation of $c_\mathrm{DIC}$ near the surface and very little near the base of the mixed layer. Conversely, the three Langmuir cases have progressively more uniform concentrations and lower variance throughout the mixed layer. Again, the more uniform vertical distribution and decreased standard deviation of the three Langmuir cases, in comparison to the non-Langmuir case, can largely be attributed to the faster vertical mixing associated with Langmuir turbulence. Figure~\ref{fig:figure6}(c) shows that the Langmuir cases all exhibit increased downward vertical flux near the surface in comparison to the non-Langmuir case. While their magnitudes are quite similar near the surface, the stronger Langmuir cases have sustained increased flux deeper into the domain.
    
\begin{figure}[t!]
\centering
\includegraphics[width=\linewidth]{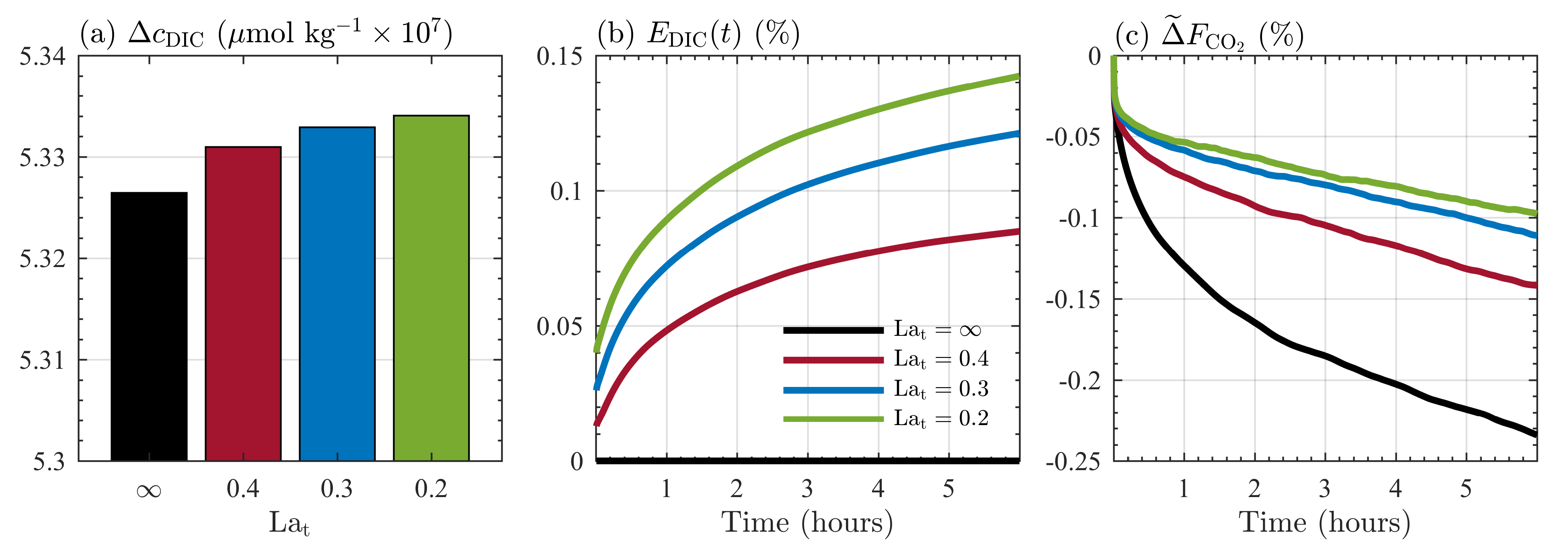}
\caption{Dependence on Langmuir number of (a) volume-integrated change in $c_\mathrm{DIC}$, where $\Delta c_\mathrm{DIC}$ is given by Eq.\eqref{cdelta}, (b) time series of the enhancement in $c_\mathrm{DIC}$ within the domain, where $E_\mathrm{DIC}$ is defined in Eq.~\eqref{percent_enhance}, and (c) time series of the normalized change in surface flux, where $\widetilde{\Delta} F_{\ce{CO2}}$ is defined in Eq.~\eqref{flux_enhance}. All panels show results for Langmuir numbers $\mathrm{La}_\mathrm{t} = \infty$, 0.4, 0.3, and 0.2 (black, red, blue, and green bars and lines, respectively) using time-dependent chemistry.}
\label{fig:figure7}
\end{figure}

As \ce{CO2} is mixed away from the surface, a larger air-sea flux results from Henry's law in Eq.~\eqref{henry}. Thus, an increase in $c_\mathrm{DIC}$ is expected to occur as the strength of Langmuir turbulence increases. Figure~\ref{fig:figure7}(a) shows the total domain-integrated change in $c_\mathrm{DIC}$ after six hours relative to the initial concentration when the air-sea flux of $\ce{CO2}$ begins (defined here to be at $t=0$). This total change, denoted $\Delta c_\mathrm{DIC}$, is expressed as
\begin{linenomath*}
\begin{equation}\label{cdelta}
\Delta c_\mathrm{DIC}(t) = \langle c_\mathrm{DIC} \rangle_V(t)-\langle c_\mathrm{DIC} \rangle_V(t=0)\,, 
\end{equation}
\end{linenomath*}
where $\langle\cdot\rangle_V$ is an average over the entire domain in $x$--$y$--$z$ directions at a particular time. Figure~\ref{fig:figure7}(a) shows that $\Delta c_\mathrm{DIC}(t=6\,\mathrm{hours})$ progressively increases as $La_\mathrm{t}$ decreases, indicating that the Langmuir cases have indeed brought additional DIC into the domain as compared to the non-Langmuir case. 

At first glance, the differences in Figure~\ref{fig:figure7}(a) may appear to be small. However, Figure~\ref{fig:figure7}(b) shows that there can be a significant enhancement in the amount of DIC brought into the mixed layer by Langmuir turbulence. This can be quantified by comparing $\Delta c_\mathrm{DIC}$ for the Langmuir and non-Langmuir cases at each time and by defining an enhancement parameter, $E_\mathrm{DIC}$, that expresses the difference relative to the non-Langmuir case. This parameter is calculated as
\begin{linenomath*}
\begin{equation}
E_\mathrm{DIC}(t) =100\times \frac{\Delta c_\mathrm{DIC}(t) - [\Delta c_\mathrm{DIC}]_\mathrm{base}(t)}{[\Delta c_\mathrm{DIC}]_\mathrm{base}(t)}\label{percent_enhance}\,,
\end{equation}
\end{linenomath*}
where $E_\mathrm{DIC}$ is expressed as a percentage and $[\Delta c_\mathrm{DIC}]_\mathrm{base}$ is the baseline change in domain-integrated $c_\mathrm{DIC}$ against which the Langmuir cases are compared. In this section, $[\Delta c_\mathrm{DIC}]_\mathrm{base}$ is taken to be $\Delta c_\mathrm{DIC}$ for the non-Langmuir case with time-dependent chemistry. Figure~\ref{fig:figure7}(b) shows that, for this simulation configuration and after six hours of constant, uniform wind and wave forcing, there is a Langmuir-induced enhancement of 0.09--0.14\% more DIC in the domain as compared to the non-Langmuir case with just wind-driven shear turbulence.
      
Fundamentally, the observed differences in new DIC brought into the domain are due to differences in the flux rate of \ce{CO2} across the air-sea interface, given by $F_{\ce{CO2}}$ in Eq.~\eqref{flux}. Figure~\ref{fig:figure7}(c) shows the change in horizontally ($x$--$y$) averaged $F_{\ce{CO2}}$ as a function of time, denoted $\widetilde{\Delta} F_{\ce{CO2}}$, where the notation $\widetilde{\Delta}$ reflects the fact that the change is normalized by the average $F_{\ce{CO2}}$ at the initial time. This quantity is calculated as
\begin{linenomath*}
\begin{equation}
\widetilde{\Delta} F_{\ce{CO2}} (t)  =100\times \frac{\langle F_{\ce{CO2}}\rangle(t) - \langle F_{\ce{CO2}}\rangle(t=0)}{\langle F_{\ce{CO2}}\rangle(t=0)}\label{flux_enhance}\,,
\end{equation}
\end{linenomath*}
where, as with $E_\mathrm{DIC}$ in Eq.~\eqref{percent_enhance}, $\widetilde{\Delta} F_{\ce{CO2}}$ is expressed as a percentage. The time series of $\widetilde{\Delta} F_{\ce{CO2}}$ in Figure~\ref{fig:figure7}(c) show that all cases have a sharp initial decline in air-sea flux rate. However, $\widetilde{\Delta} F_{\ce{CO2}}$ for the non-Langmuir case continues to decrease at a faster rate in comparison to the three Langmuir cases, indicating that Langmuir flux enhancement may persist over diurnal and synoptic time scales if saturation does not occur. If there is a build-up of \ce{CO2} at the surface, the air-sea gradient in \ce{CO2} concentration decreases, thereby decreasing $F_{\ce{CO2}}$ locally, and $\langle F_{\ce{CO2}}\rangle$ over the entire surface. If, instead, the concentration of \ce{CO2} at the surface is maintained at a lower value for a longer period of time, the air-sea gradient in \ce{CO2} is relatively unchanged and $F_{\ce{CO2}}$ is not reduced as dramatically. 

\subsection{Effects of Chemical Model Fidelity on Carbonate Chemistry\label{chemistry effect}}
To isolate the effect of chemical model fidelity on the air-sea flux rate of \ce{CO2} and the resulting enhancement of DIC within the upper ocean, this section compares the shear-only non-Langmuir case and the $\text{La}_\mathrm{t} = 0.3$ case for each of the three chemistry models (i.e., TC, EC, and NC).

\begin{figure}[b!]
\centering
\includegraphics[width=\linewidth]{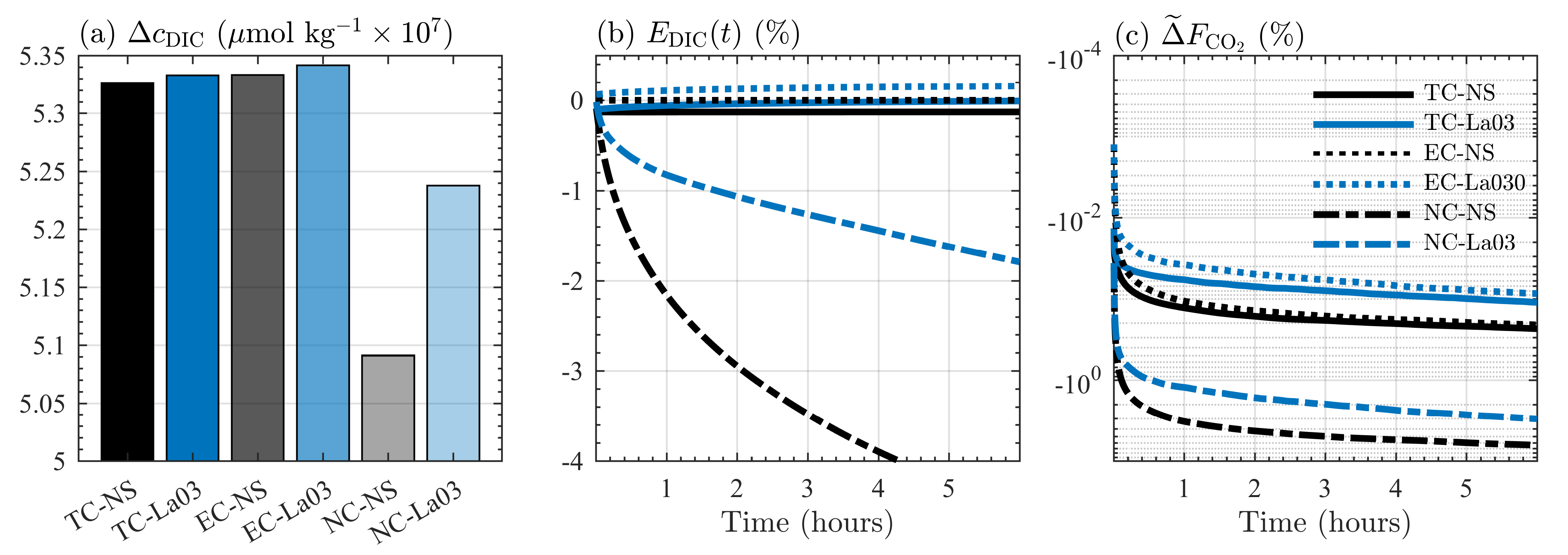}
\caption{Dependence on chemical model fidelity of (a) volume-integrated change in $c_\mathrm{DIC}$, where $\Delta c_\mathrm{DIC}$ is given by Eq.\eqref{cdelta}, (b) time series of the enhancement in $c_\mathrm{DIC}$ within the domain, where $E_\mathrm{DIC}$ is defined in Eq.~\eqref{percent_enhance} and the non-Langmuir equilibrium chemistry case is used as a baseline, and (c) time series of the normalized change in surface flux, where $\widetilde{\Delta} F_{\ce{CO2}}$ is defined in Eq.~\eqref{flux_enhance}. All panels show results for the time-dependent (TC), equilibrium (EC), and no (NC) chemistry models, for both non-Langmuir (NS) and $\text{La}_\mathrm{t}$ = 0.3 (La03) cases.}
\label{fig:figure8}
\end{figure}

The volume-integrated change in DIC [defined in Eq.~\eqref{cdelta}] shown in Figure~\ref{fig:figure8}(a) indicates that both the TC and EC models bring more carbon into the domain in comparison with the NC model. This is because carbonate chemistry, in either time-dependent or equilibrium forms, provides a sink of \ce{CO2} and preserves the air-sea gradient, resulting in chemistry flux enhancement. The EC case exceeds the TC case in carbon uptake. This occurs because reactions in the EC case are infinitely fast and thus aqueous \ce{CO2} is instantly converted into its respective proportions (based on the local temperature, salinity, DIC, and alkalinity) of \ce{CO2}, \ce{HCO3^-}, and \ce{CO3^2-}. For the TC case, by contrast, \ce{CO2} persists for a finite amount of time before reacting and/or being removed by advection, leaving an increased surface concentration of \ce{CO2}, which slows fluxes. Notably, the same trend between the non-Langmuir case and $\text{La}_\mathrm{t} = 0.3$ is seen for each chemistry model, however, the difference between the two sets varies with the chemistry model (combined effects are discussed in the next section).
     
Figure~\ref{fig:figure8}(b) shows the percent enhancement in the volume-integrated new DIC as a function of time for each of the chemistry models with respect to the non-Langmuir EC case. The enhancement is given by Eq.~\eqref{percent_enhance}, with $[\Delta c_\mathrm{DIC}]_\mathrm{base}$ now defined as the non-Langmuir EC case. This case is chosen as the baseline since this chemistry model and physical configuration resemble those used in ESMs. Once again, the general trends between the non-Langmuir and $\text{La}_\mathrm{t} = 0.3$ cases are consistent across the different chemistry models, with $\text{La}_\mathrm{t} = 0.3$ showing greater enhancement, but the detailed differences between these cases are dependent on the chemistry model (i.e., 0.12\%, 0.16\%, and 2.7\% increases in new DIC between the non-Langmuir and $\text{La}_\mathrm{t} = 0.3$ cases for the TC, EC, and NC models, respectively). 
          
Finally, Figure~\ref{fig:figure8}(c) shows time series of the change in air-sea flux rate of \ce{CO2} [defined in Eq.~\eqref{flux_enhance}] for the three chemistry models. The NC case undergoes a dramatic decline in air-sea flux rate as non-reactive aqueous \ce{CO2} builds up and slows fluxes. The two reactive cases (TC and EC), by contrast, have much higher flux rates. Both the TC and EC cases convert \ce{CO2} into \ce{HCO3^-} and \ce{CO3^2-}, which maintains a greater air-sea gradient of \ce{CO2} and allows more \ce{CO2} to enter the ocean. Comparing the two reactive cases, the EC case has an elevated flux rate over the TC case due to its faster reaction sink. 

\subsection{Combined Effects of Langmuir Turbulence and Chemical Model Fidelity}
The previous sections have shown that both the enhanced vertical flux due to Langmuir turbulence and the chemical model fidelity affect the air-sea flux rate and reduce surface concentrations of \ce{CO2}, thereby impacting the DIC content of the oceanic mixed layer. Here, the combined effects of Langmuir turbulence enhancement and chemical model fidelity are considered. 

\begin{figure}[b!]
\centering
\includegraphics[width=\linewidth]{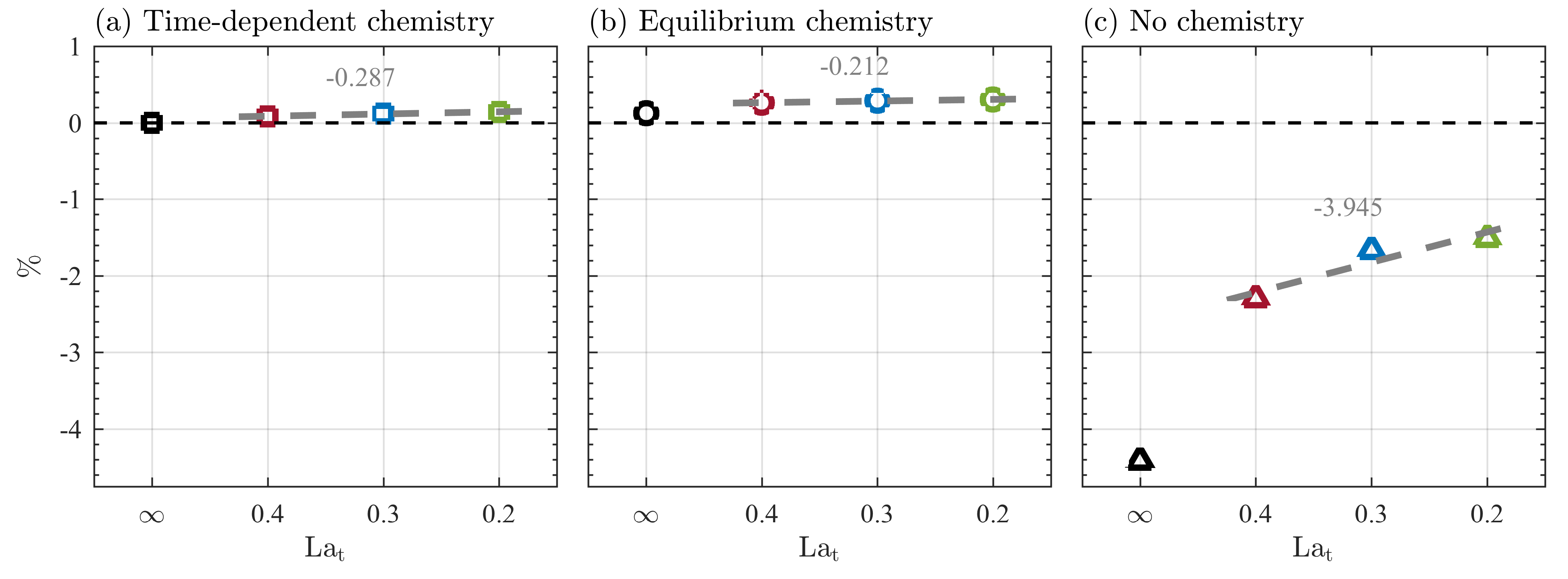}
\caption{Percent enhancement of new DIC, denoted $E_\mathrm{DIC}$ and defined in Eq.~\eqref{percent_enhance}, in comparison to the non-Langmuir, time-dependent chemistry case for the three chemical models (a--c) and the four Langmuir cases. Gray dashed lines are least-squares fits for the decrease in percent enhancement as a function of Langmuir number and the gray number is the slope of the fit. All panels show results for Langmuir numbers $\mathrm{La}_\mathrm{t} = \infty$, 0.4, 0.3, and 0.2 (black, red, blue, and green symbols, respectively).}
\label{fig:figure9}
\end{figure}

Figure~\ref{fig:figure9} shows the percent enhancement in the volume-integrated new DIC at six hours for each of the three chemistry models and for each of the four Langmuir cases. The expression in Eq.~(\ref{percent_enhance}) is once again used to compute the percentage of new DIC, and all values are now referenced to the non-Langmuir, time-dependent case, which occupies the zero value in panel (a) of Figure~\ref{fig:figure9}. 
     
Figure~\ref{fig:figure9} shows that the Langmuir flux enhancement is a function of the chemistry model, reflecting a complex, non-linear relationship between the chemical model and small-scale turbulence. Langmuir turbulence in the NC case provides a large enhancement over just wind-driven shear turbulence, while TC and EC cases have modest Langmuir enhancement. Consequently, the effect of Langmuir turbulence on air-sea fluxes of a gas varies substantially depending on whether the gas is reactive (e.g., \ce{CO2}), or non-reactive (e.g., oxygen).
     
Smaller differences distinguish the two reactive cases. For the EC case, the Langmuir enhancement is greater than in the TC case, yet successive increases in Langmuir strength do not affect the EC case as much as in the TC case.
    
\section{Discussion\label{sec:discuss}}
In the following, the implications of the results described in Section \ref{results} are discussed with respect to ESMs, and variations in these results for different ocean conditions are outlined. The latter discussion is focused, in particular, on how the strength of the interactions between vertical mixing and chemical processes vary over the global ocean, as well as over diurnal and seasonal cycles. 

\subsection{Implications for Earth System Models}
The results in Section \ref{langmuir effect} indicate that over the entire ocean surface, approximately \SIrange{0.07}{0.1}{\peta\gram} of extra carbon per year is brought into the ocean due to the presence of Langmuir turbulence. This estimate is based on an approximate global air-to-sea \ce{CO2} flux for 2000--2009 of \SI{80}{\peta\gram\carbon\per\year}; for reference, the estimated global net air-sea \ce{CO2} flux for this same period is \SI{2.3 \pm 0.7}{\peta\gram\carbon\per\year} \citep{ciais2013}. This assumes that these exact conditions remain constant throughout the year across the entire ocean surface and that the air concentration of \ce{CO2} is always 10\% greater than the mixed layer equilibrium concentration. Regional and seasonal deviations are likely, as imbalances will occur, for example, during upwelling, cooling, and warming events. These deviations are not random and may introduce systematic biases depending on the turbulence forcing mechanism. Nevertheless, the results in Section \ref{langmuir effect} indicate that Langmuir turbulence has a meaningful effect on the uptake of carbon by the ocean.

Similarly, the results in Section \ref{chemistry effect} indicate that the finite-time delay in \ce{CO2} conversion due to the use of TC chemistry would result in a roughly \SI{0.1}{\peta\gram} decrease in the global uptake of carbon by the ocean in comparison with models that use the instantaneous EC chemistry. Even larger discrepancies in total domain carbon are found between the NC and other cases.
     
These changes in \ce{CO2} flux rate with chemical model and Langmuir turbulence are on the same order as basin-scale differences in flux rate for different ESMs. Most current ESM simulations make two assumptions: (\emph{i}) that boundary layer turbulence effects on all chemical species can be parameterized in the same way, and (\emph{ii}) that carbonate chemistry is virtually instantaneous in comparison to turbulent processes, and thus can be represented by an equilibrium chemistry model. Results from the present study thus contradict these assumptions and indicate that errors from both assumptions combine in a complex and non-linear way. It should be noted, however, that errors resulting from the neglect of Langmuir turbulence and the use of equilibrium chemistry are likely to be dominated by errors in other physical models within ESMs, particularly for globally integrated annual quantities. The greatest impacts from the inclusion of a Langmuir parameterization and finite-rate chemistry are thus likely to be felt at regional spatial scales and over shorter time scales.  

\subsection{Dependence on Ocean Conditions\label{time}}
The physical parameters chosen for the simulations, as outlined in Section \ref{physics} and summarized in Table \ref{tab:setup}, were selected on the basis of convenience combined with realism. These parameters result in potentially matched timescales for chemical and turbulent processes, and the extent of the matching can be expressed using various non-dimensional timescale ratios. To this end, the Damk\"{o}hler number is given as the ratio of the turbulent advection timescale, $\tau_\mathrm{t}$, to a characteristic timescale of the overall reaction process, $\tau_\mathrm{c}$, namely $\mathrm{Da} = \tau_\mathrm{t}/\tau_\mathrm{c}$.
    
In general terms, $\tau_\mathrm{t}$ can be estimated from the integral or eddy turnover timescale. For a configuration similar to that studied here, \citet{teixeira2010} estimated a near-surface integral timescale of \SI{430}{\second} for a Langmuir turbulence simulation with $u_\tau = \SI{6.1e-3}{\meter \per \second}$. Calculations of the integral timescale in the present simulation yield a similar result, and so $\tau_\mathrm{t}$ can be estimated as $\tau_\mathrm{t} \approx \SI{400}{\second}$. In determining $\tau_\mathrm{c}$, it is common to use a characteristic timescale associated with the global rate of reaction, and the relaxation time after a 10\% perturbation to the concentration of \ce{CO2} (discussed in more detail in Section \ref{reactions}) gives $\tau_\mathrm{c} \approx \SI{60}{\second}$. As a result, Da in the present case can be estimated as $\mathrm{Da} \approx \SI{400}{\second} / \SI{60}{\second} = 6.7$. This value of Da indicates that interactions between Langmuir turbulence and carbonate chemical reactions are important, but that reactions are favored. This is consistent with results outlined in Section \ref{results}, particularly with respect to those in Figure \ref{fig:figure9} where variations in the chemical model fidelity were shown to have a larger impact on the amount of DIC in the mixed layer than the strength of Langmuir turbulence, at least for the present ocean conditions. 
     
Variations in the value of Da can be inferred for different ocean conditions from the analysis of \citet{teixeira2010}. In that study, $\tau_\mathrm{t} = \SI{430}{\second}$ was obtained from the turbulent kinetic energy $k$ and the dissipation rate $\varepsilon$ following the $k-\varepsilon$ modeling approach, which in later work was extended to Langmuir turbulence in a variety of settings \citep{grant2009,belcher2012}. This approach, validated against LES, predicts $\tau_\mathrm{t} \sim k/\varepsilon$, where
\begin{linenomath*}
\begin{equation}
k\propto u_\tau^2 \quad \mathrm{or} \quad [u_\tau^2 u_\mathrm{s}(0)]^{2/3} \quad \mathrm{or} \quad (B_0 h_\mathrm{b})^{2/3} \,,
\end{equation}
\end{linenomath*}
for wind-, wave-, and convection-dominated conditions, respectively, and
\begin{linenomath*}
\begin{equation}
\varepsilon\propto \frac{2u_\tau^3[1-\exp(-\mathrm{La}_\mathrm{t}/2)]}{h_\mathrm{b}} + 0.22 \frac{u_\tau^2 u_\mathrm{s}(0)}{h_\mathrm{b}} + 0.3 B_0\,.
\end{equation}
\end{linenomath*}
Here $B_0$ is the buoyancy flux and $h_\mathrm{b}$ is the turbulent boundary layer depth. Thus, for wind-dominated conditions $\mathrm{Da}\propto h_\mathrm{b} / [u_\tau (1 - \exp(-\mathrm{La}_\mathrm{t}/2))]$, for wave-dominated conditions $\mathrm{Da}\propto h_\mathrm{b} / [u_\tau^2 u_\mathrm{s}(0)]^{1/3}$, and for convection-dominated conditions $\mathrm{Da}\propto (h_\mathrm{b}^2/B_0)^{1/3}$.  Observations suggest that the real ocean is typically somewhere between these different scalings \citep{belcher2012,li2017}.  

Assuming that $h_b$ is proportional to the initial mixed layer depth $H_{\text{ML},0}$, the scalings above indicate that $\mathrm{Da} \propto H_{\text{ML},0}$ or $H_{\text{ML},0}^{2/3}$. In the present simulations, $H_{\text{ML},0} = \SI{30}{\meter}$, but variations between \SI{10}{\meter} and \SI{500}{\meter} can occur in the real ocean depending on location and season \citep{li2016}. The depth $h_b$ has a similar range, but a value of \SI{30}{\meter} or less is typical in the tropics and during the summertime. This range roughly corresponds to a decrease in Da by a factor of 3 for the shallowest layers, or 15 times larger for the deepest layers. In the former case, this corresponds to a stronger interaction between Langmuir turbulence and carbonate chemistry, while the latter case corresponds to a weaker interaction. Assuming a fixed reaction rate (i.e., neglecting temperature and salinity effects on reaction rates) and using the \citet{large2009} monthly-mean wind stresses and the updated \citet{de2004} mixed layer depth climatology, $\mathrm{Da}$ based on wind stress scaling was estimated to have 90\% confidence limits of 3 and 20 with a median near 7.  As higher-frequency winds tend to induce faster mixing during intermittent events, it is expected that these estimates are biased toward high Da.
     
Surface cooling is not used here, but $B_0$ is an important scaling parameter for the amount of convective mixing in the world oceans, and varies from \SI{1e-9}{\meter^2 \second^{-3}} to \SI{5e-7}{\meter^2 \second^{-3}} by season and time of day. As a result, the dependence on $B_0$ might further decrease Da by a factor of 7 under extreme events, once again corresponding to stronger turbulence-chemistry interactions. Similarly, the typical wind stress over the ocean is \SI{0.1}{\newton \meter^{-2}}, or four times the value used here, which would roughly halve the value of Da (holding other parameters fixed). Using the \citet{large2009} monthly-mean buoyancy fluxes and the updated \citet{de2004} mixed layer depth climatology, the convection-based $\mathrm{Da}$ scaling using the same datasets had a factor of 7 spread for the 90\% confidence limits, although a median value cannot be calculated without the normalization factor from a convectively-forced LES.
     
Note that it is also possible to define and estimate other relevant Damk\"ohler numbers. For example, a method to determine the smallest possible timescale of oceanic turbulence is to use the \citet{kolmogorov1941} time scale, typifying the timescale of the smallest-scale turbulence in the ocean.  However, the vast range of energy dissipation rates throughout the world ocean does not make this estimate very precise \citep{pearson2018} beyond a range such as \SI{0.1}{\second} to \SI{1000}{\second}, which spans a wide range enclosing $\mathrm{Da}={\cal O}(1)$. Similarly, the second Damk\"ohler number, ${\rm Da}_2$, is the dimensionless ratio of mass diffusion timescale $\tau_\kappa$ to $\tau_\mathrm{c}$, namely
\begin{linenomath*}
\begin{equation}
{\rm Da}_2=\frac{\tau_\kappa}{\tau_c}\,.
\end{equation}
\end{linenomath*}
In the models used here, the subgrid-scale viscosity, buoyancy diffusivity, and tracer diffusivity are spatially-varying according to the scheme proposed by \citet{sullivan1994}.  The diffusivities provided by this scheme are much larger than the molecular values for seawater.  However, they are scaled in a flow-aware way through the LES approach so that $\mathrm{Da}\ll {\rm Da}_2$. As the simulations and Langmuir turbulence in the ocean are already in the regime where $\mathrm{Da}\ge 1$, it is expected that the consequences of using the LES diffusivities rather than the molecular diffusivities will be small since ${\rm Da}_2\gg \mathrm{Da}1\ge 1$. Note that the modeled LES diffusivities are not used for the Schmidt number when calculating the piston velocity from Eq.~\eqref{schmidt} \citep{wanninkhof1992}.  Instead, the values of seawater Schmidt number consistent with observations are used, which probably includes the effects of turbulence not resolved in these simulations.
     
In summary, the wind stress and surface cooling used here are conservative estimates of conditions typically observed in the ocean, while boundary layer depth is typical of the tropics or midlatitude summers. From the estimate $\mathrm{Da} \approx 6.7$, lower values of Da will result for stronger surface winds and cooling, while larger Da will result from deeper boundary layers. Thus, finite-time chemistry effects will typically be strongest during mixed layer deepening under strong wind and cooling events, such as cold air outbreaks. A full assessment of the climatology of Da for a full range of seasonal and regional conditions is beyond the scope here, but is planned for future work.

\section{Conclusions}
The interactions between carbonate chemical reactions and turbulent mixing in the upper ocean have been examined using LES for four different strengths of wave forcing and three different carbonate chemistry models, from infinitely slow non-reactive chemistry to infinitely fast equilibrium chemistry. The novel model in between is a time-dependent seven-species carbonate chemistry model that uses a QSS assumption for \ce{H^+} and is integrated using an efficient RKC solver that is robust for stiff problems.

The results presented here indicate that enhanced vertical mixing by Langmuir turbulence results in a small, but measurable, increase in DIC in the ocean mixed layer as compared to a case with no Langmuir turbulence. Conversely, the use of an equilibrium chemical model results in a small, but measurable, reduction of DIC in the mixed layer as compared to a more realistic time-dependent model. The combined effects of Langmuir turbulence and chemical model fidelity are complicated and coupled, but the effects of Langmuir turbulence are more pronounced when using time-dependent chemistry than when using equilibrium chemistry. 

With respect to ESMs, this study has resulted in three major insights. First, compared with shear-only turbulence, Langmuir turbulence increases the flux rate of \ce{CO2} across the air-sea interface by approximately 0.1\%, or \SIrange{0.07}{0.1}{\peta\gram\carbon\per\year} globally. Second, the more accurate finite-time chemistry decreases the flux rate of \ce{CO2} into the domain by approximately 0.1\%, or \SI{0.1}{\peta\gram\carbon\per\year}, in comparison with equilibrium chemistry (and increases versus no chemistry). Third, Langmuir turbulence has a much greater effect on flux rates of a non-reactive gas such as oxygen than on a reactive gas such as \ce{CO2}. The magnitude of these differences is also expected to depend on other aspects of the ocean state, requiring further study in the future. 
     
In the future, additional research is required to determine whether Langmuir turbulence and finite-rate chemistry have different impacts at other ocean locations and for different conditions. The effects of wave breaking and bubbles are also likely to be important in air-sea fluxes of \ce{CO2}, and future simulations are planned using a bubble parameterization. Finally, carbonate chemistry evolution was only examined over a relatively short period in the present study, and longer simulations that incorporate diurnal and seasonal cycles will be performed in the future, thus providing more accurate estimates of the annual impacts of Langmuir turbulence and chemical model fidelity on \ce{CO2} flux rates.

\acknowledgments
KMS, PEH, and NSL were supported by NSF OCE-1258995, KEN was supported by NSF OAC-1535065, and BFK was supported by NSF OCE-1258907 and a grant from The Gulf of Mexico Research Initiative. The data analyzed in this paper are available from Mendeley Data (https://data.mendeley.com). This work utilized the RMACC Summit supercomputer supported by NSF (ACI-1532235, ACI-1532236), CU-Boulder, and CSU, as well as the Yellowstone (ark:/85065/d7wd3xhc) and Cheyenne (doi:10.5065/D6RX99HX) supercomputers provided by NCAR CISL, sponsored by NSF.

\bibliography{refs.bib}

\end{document}